\documentclass[lettersize,journal]{IEEEtran}
\usepackage{amsmath,amsfonts,amssymb}
\usepackage{amsthm}
\newtheorem{proposition}{Proposition}
\usepackage{algorithmic}
\usepackage{algorithm}
\usepackage{array}
\usepackage[caption=false,font=footnotesize,labelfont=rm,textfont=rm]{subfig}
\usepackage{textcomp}
\usepackage{stfloats}
\usepackage{url}
\usepackage{verbatim}
\usepackage{graphicx}
\usepackage{xcolor}
\usepackage{placeins}
\usepackage{cite}
\begin{document}

\title{Cross-Region Cooperation of UAV Swarms Towards Air-Ground ISAC via
Hierarchical Multi-Agent Reinforcement Learning}

\author{Linghui Miao,~\IEEEmembership{Graduate Student Member,~IEEE}, Nhan Thanh Nguyen,~\IEEEmembership{Senior Member,~IEEE}, and Shijian Gao,~\IEEEmembership{Member,~IEEE}
\thanks{Part of this work has been presented at the IEEE/CIC International Conference on Communications in China (ICCC) 2026~\cite{miao2026uavswarm}.}}

\markboth{Journal of \LaTeX\ Class Files,~Vol.~14, No.~8, August~2021}%
{Shell \MakeLowercase{\textit{et al.}}: A Sample Article Using IEEEtran.cls for IEEE Journals}

\IEEEpubid{0000--0000/00\$00.00~\copyright~2021 IEEE}

\maketitle

\begin{abstract}
\boldmath
The limited communication resources and sensing perspectives of a single uncrewed aerial vehicle (UAV) motivate UAV swarms for integrated sensing and communication (ISAC) in air-ground networks. However, in multi-region scenarios, regional UAV allocations may not match vehicle demand, while local observations may lack sufficient diversity for accurate sensing. To address these issues, this paper proposes a cross-region cooperation framework for UAV swarms. For communication, cross-region UAV reallocation reduces service disparities by matching resources to vehicle demand. For sensing, combining UAV observations from different regions increases observation diversity, while adaptive handshaking guided by marginal Cram\'er--Rao bound (CRB) utilities balances phase synchronization gains against handshaking energy consumption. Joint optimization of UAV positions and transmit power within each region then further improves communication service and sensing performance. These coupled decisions are coordinated by a hierarchical multi-agent reinforcement learning (H-MARL) approach through slow-timescale inter-region UAV reallocation and fast-timescale intra-region UAV control. Under the considered simulation settings, reallocation can improve quality of service (QoS) by over 20 percentage points compared with retaining the initial allocation under severe resource-demand mismatches. Cooperative sensing with adaptive handshaking incurs only about a $5.0\%$ increase in mean CRB relative to perfect synchronization.
\end{abstract}

\begin{IEEEkeywords}
Air-ground networks, integrated sensing and communication, multi-agent reinforcement learning, dual-timescale optimization, cross-region cooperation.
\end{IEEEkeywords}

\section{Introduction}

\IEEEPARstart{T}{o} support communication and sensing services in the volumetric air-ground space, uncrewed aerial vehicles (UAVs) are emerging as key aerial nodes owing to their flexible deployment, controllable three-dimensional mobility, and favorable line-of-sight air-ground links \cite{gao2026iscc}. These features are well aligned with integrated sensing and communication (ISAC), supporting data transmission and target sensing over shared wireless resources \cite{liu2022isac}. Therefore, UAV-enabled ISAC is a promising paradigm for low-altitude and mobile networks~\cite{cooperative2026lae}. Realizing this potential requires coordinating platform positions and transmission resources as service demands and operating conditions vary over time and across locations. Such coordination helps aerial platforms adapt to these variations while supporting reliable communication and accurate sensing.

Research on single-UAV ISAC has investigated trajectory and beamforming optimization \cite{he2025energyisac}, as well as resource allocation to coordinate communication and sensing performance \cite{uavnoma2026isac,uavris2026marlisac}. However, the coverage, onboard resources, and observation geometry of an individual UAV constrain the scale and flexibility of its services. UAV swarms expand the available communication resources and enable sensing from multiple locations. Exploiting these capabilities requires cooperation among UAVs, whose movements and transmissions jointly affect communication and sensing performance \cite{qindrl2023uavisac,cooptraj2026resource}. Existing studies have coupled UAV trajectory design with user association and beamforming to coordinate aerial deployment and ground-user service \cite{jointuav2026trajassocbeam}. UAV deployment and sensing-communication resources have also been jointly optimized for federated edge learning \cite{tang2025iscc}. Beyond mobility and transmission design, collaborative sensing and communication introduce task allocation decisions, which have been jointly optimized with transmit power \cite{multiuav2026taskpower}. The spatial distribution of ground users further requires adapting UAV deployment across geographic service regions. Density-based vehicle clustering identifies and updates congested regions, where UAV clusters are deployed to support continuous vehicular task offloading \cite{liao2024adaptive}. These studies highlight the importance of coordinating aerial resources across service regions for air-ground ISAC.
\IEEEpubidadjcol

From the communication perspective, synesthesia of machines (SoM) supports adaptive ISAC transmission using positioning and channel information in vehicular networks \cite{yang2025som} and visual and radio-frequency information in air-ground networks \cite{yang2026somairground}. Beyond transmission control, hierarchical task offloading and cluster-based scheduling accommodate time-varying workloads \cite{hvtask2026offloading,cluster2026sagin}, while UAV cluster reconfiguration addresses spatial service adaptation \cite{cluster2026uavmobility}. These studies highlight two requirements for UAV-swarm communication: \emph{demand-aware allocation of aerial resources} and \emph{adaptive mobility and transmission control}. Reallocating UAVs across regions can better align aerial resources with vehicle demand, but communication performance also depends on mobility and transmission control within each region. From the sensing perspective, network-level ISAC exploits coordination among geographically separated sensing nodes to improve sensing accuracy \cite{networklevel2026isac}. Fusing uplink and downlink observations improves target angle estimation \cite{chowdary2024hybrid}. For UAV swarms, cooperative perception and trajectory optimization exploit UAV mobility to improve sensing performance \cite{multiuav2026perception}. Beyond observation geometry, synchronization and beam management help maintain distributed sensing gains \cite{coopsensing2026challenges}. These findings identify two requirements for UAV-swarm sensing: \emph{diverse observation geometries} and \emph{accurate inter-UAV synchronization}. Extending sensing across regions can increase observation diversity, but it requires reducing phase synchronization errors across regions.

Joint optimization of UAV trajectories and transmission resources considers both communication and sensing performance in UAV-enabled ISAC \cite{he2025energyisac}. As network conditions change, repeated model-based optimization can impose a substantial computational burden on online control. To address this computational burden, deep reinforcement learning (DRL) has been used to learn UAV trajectory and resource allocation policies through environment interaction \cite{qindrl2023uavisac}. In multi-UAV settings, multi-agent reinforcement learning (MARL) supports joint trajectory and beamforming design \cite{gao2024marl}, as well as adaptive control for UAV-enabled vehicular ISAC \cite{wang2025novel}. Hierarchical reinforcement learning coordinates trajectory planning and resource allocation for UAV-assisted data collection \cite{yuan2025hierarchical}. Graph-based learning incorporates network relationships into multi-UAV trajectory design and resource assignment \cite{gat2026uavcomm}. Together, these studies demonstrate the potential of learning-based coordination for adapting UAV decisions to changing network conditions while jointly considering communication and sensing performance.

Spatially clustered vehicle demand motivates partitioning the service area around traffic hotspots for localized UAV service and control. Cross-region cooperation addresses regional resource-demand mismatches and limited local sensing diversity, but introduces two technical challenges. First, unlike independent operation with fixed regional UAV assignments, cross-region reallocation requires deciding which UAVs to dispatch and where. This decision must balance communication gains in receiving regions against communication losses in source regions, sensing degradation, and transit energy, with these tradeoffs depending on subsequent intra-region mobility and power control. Second, neighboring regions contribute unequally to target sensing, while residual inter-region phase errors impair coherent processing. Synchronization effort must therefore reflect regional sensing contributions while balancing sensing accuracy against handshaking energy consumption. To address these challenges, this paper proposes a cross-region cooperation framework for UAV swarms towards air-ground ISAC. UAV reallocation balances regional communication service subject to a sensing-loss tolerance, while adaptive handshaking uses regional marginal Cram\'er--Rao bound (CRB) utilities to coordinate synchronization. A hierarchical multi-agent reinforcement learning (H-MARL) approach coordinates slow-timescale reallocation and fast-timescale mobility, transmit-power, and handshaking control. The main contributions are summarized as follows:
\begin{itemize}
    \item A cross-region cooperation framework for UAV swarms towards air-ground ISAC is proposed. A joint optimization problem is formulated over regional UAV allocation, UAV mobility, transmit power, and inter-region synchronization, capturing their coupled effects on communication service, sensing accuracy, and energy consumption.

    \item A UAV reallocation mechanism is developed to improve communication quality of service (QoS) and regional fairness. It jointly selects neighboring destinations and dispatched UAVs, subject to a CRB-based sensing-loss tolerance for each selected UAV.
    
    \item An adaptive handshaking mechanism is developed for cross-region cooperative sensing. Marginal CRB utilities quantify regional sensing contributions and guide the aggregation of UAV-level handshaking preferences into a common frequency.

    \item The proposed H-MARL approach is designed to coordinate regional reallocation and local UAV control across two timescales. A regional graph policy captures resource and demand conditions together with neighboring-region relationships, while local UAV policies adapt their operation to the updated regional allocation.
\end{itemize}

The remainder of this paper is organized as follows. Section~II presents the system model of the air-ground ISAC network. Section~III defines the communication, sensing, and energy utilities. Section~IV formulates the joint problem and its two hierarchical subproblems. Section~V presents the H-MARL framework and policy optimization. Section~VI presents the simulations, and Section~VII concludes the paper.

\emph{Notation:} Boldface lowercase and uppercase letters denote vectors and matrices, respectively. The superscripts $T$ and $-1$ denote transpose and matrix inverse. $\|\cdot\|$ denotes the Euclidean norm, while $|\cdot|$ denotes absolute value for scalars and cardinality for sets. The notation $\mathcal N_{\bar m}\triangleq\mathcal N_m\setminus\{m\}$ denotes the set obtained by excluding $m$ from $\mathcal N_m$. The operators $\operatorname{Tr}(\cdot)$, $\operatorname{diag}\{\cdot\}$, and $\mathbb{E}[\cdot]$ denote trace, diagonal matrix construction, and expectation, respectively. $\mathbf{1}_{\{\cdot\}}$ denotes the indicator function. For symmetric matrices, $\mathbf{A}\succeq\mathbf{B}$ ($\mathbf{A}\succ\mathbf{B}$) means that $\mathbf{A}-\mathbf{B}$ is positive semidefinite (positive definite).

\section{System Model}
\label{sec:system_model}
This section presents the air-ground network structure and the two decision timescales, followed by the implementation of communication and sensing through UAV swarm coordination.

\subsection{Air-Ground Network}

\begin{figure}[!t]
\centering
\includegraphics[width=\columnwidth]{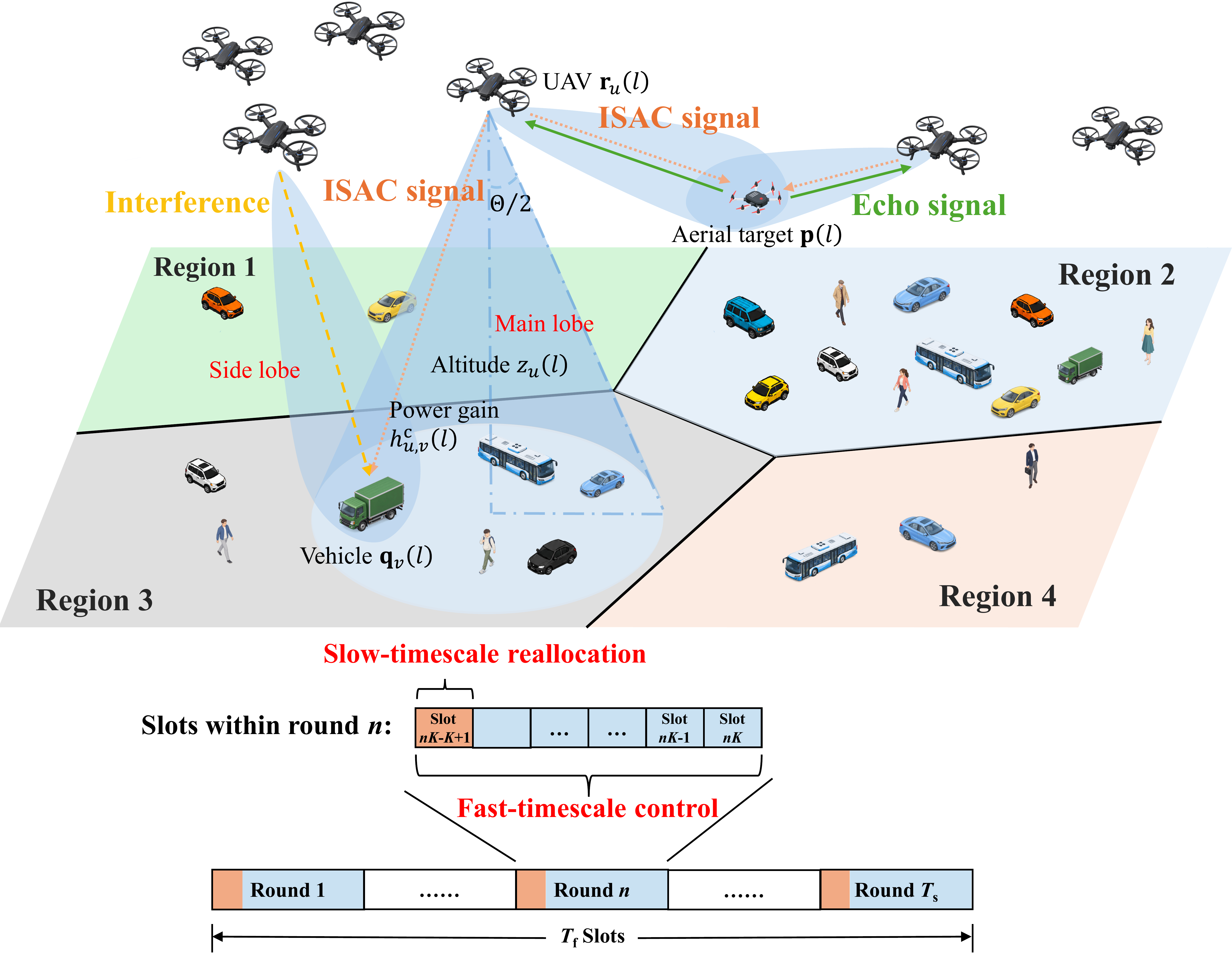}
\caption{Air-ground ISAC system with cross-region cooperation of UAV swarms.}
\label{fig:system_model}
\end{figure}

As illustrated in Fig.~\ref{fig:system_model}, we consider an air-ground ISAC system in which a UAV swarm serves ground vehicles and cooperatively senses an aerial target. The service area is partitioned into $M$ non-overlapping Voronoi regions centered at traffic hotspots. Let $\mathcal M=\{1,\ldots,M\}$ denote the region set and $\mathbf c_m$ the center of region $m\in\mathcal M$. For each region $m$, let $\mathcal N_m\subseteq\mathcal M$ denote the set consisting of region $m$ and its neighboring regions, while $\mathcal N_{\bar m}$ contains only its neighboring regions. Only interactions within a region or between neighboring regions are considered. The system operates over two decision timescales on a common slotted timeline. Fast-timescale slots are indexed by $l\in\mathcal T_{\mathrm f}=\{1,\ldots,T_{\mathrm f}\}$, each with duration $\Delta t$. Cross-region UAV reallocation is performed once every $K$ slots, with $n\in\mathcal T_{\mathrm s}=\{1,\ldots,T_{\mathrm s}\}$ indexing the reallocation rounds and $T_{\mathrm f}=KT_{\mathrm s}$. The reallocation decision for round $n$ is made at the beginning of slot $l=(n-1)K+1$. Between reallocation decisions, UAV mobility, transmit power, and handshaking preferences are adjusted at every fast-timescale slot. In each time slot of duration \(\Delta t\), the UAV and vehicle positions are first updated according to their motion. The slot duration is assumed sufficiently short for the updated spatial configuration to be treated as quasi-static when evaluating communication and sensing performance.

Let $\mathcal U=\{1,\ldots,U\}$ and $\mathcal V=\{1,\ldots,V\}$ denote the sets of UAVs and ground vehicles, respectively. At fast-timescale slot $l$, $\mathcal U_m(l)\subseteq\mathcal U$ is the set of UAVs currently available in region $m$, and $\mathcal V_m(l)\subseteq\mathcal V$ is the set of vehicles located in that region. The positions of UAV $u$, vehicle $v$, and the aerial target are denoted by $\mathbf r_u(l)=[x_u(l),y_u(l),z_u(l)]^T$, $\mathbf q_v(l)=[x_v(l),y_v(l),0]^T$, and $\mathbf p(l)=[x_p(l),y_p(l),z_p(l)]^T$, respectively. The corresponding UAV interaction set is $\mathcal C_m(l)=\bigcup_{i\in\mathcal N_m}\mathcal U_i(l)$.

\subsection{ISAC Implementation with UAV Swarms}

Both main-lobe and side-lobe gains are considered in communication and sensing. For communication, UAVs in $\mathcal U_m(l)$ provide downlink service to vehicles in $\mathcal V_m(l)$, with vehicle association accounting for link quality and bandwidth sharing. Each UAV equally divides its bandwidth and transmit power among served vehicles using orthogonal frequency resources, while bandwidth reuse among UAVs causes intra-region and inter-region interference. A vehicle is successfully served when its achievable rate meets the minimum requirement. Cross-region UAV reallocation matches regional resources to vehicle demand to reduce service disparities. For sensing, when the aerial target's horizontal projection lies in region $m$, UAVs in $\mathcal C_m(l)$ combine observations across regions to increase observation diversity. Mutually orthogonal waveforms allow transmit-receive channels to be separated for coherent processing and target localization. We consider an operating regime with sufficiently accurate intra-region synchronization and neglect its residual errors to focus on synchronization uncertainty between regional references. Inter-region handshaking reduces residual timing, frequency, and phase offsets between these references. Adaptive handshaking combines UAV-level preferences into a common frequency according to regional sensing contributions, balancing synchronization accuracy against energy consumption.

At each reallocation round, UAVs may be redistributed among neighboring regions to improve communication service while limiting sensing degradation. Under the resulting allocation, UAV mobility, transmit power, and handshaking preferences are adjusted at each fast-timescale slot. Section~III quantifies communication, sensing, and energy utilities.

\section{Characterization of Utility Functions for UAV-Swarm-Enabled ISAC}
\label{sec:utility_functions}
In this section, we first quantify communication service utility, then characterize cross-region cooperative sensing utility, and finally model energy consumption for propulsion, ISAC transmission, and inter-region handshaking.

\subsection{Cross-Region Communication Service Utility}
Each UAV employs a downward directional antenna with half-power beamwidth $\Theta$. Let $\psi_{u,v}(l)$ denote the angle between UAV $u$'s antenna boresight and its direction toward vehicle $v$. Following the sectored antenna model in~\cite{Bai2015Coverage}, the antenna gain is
\begin{equation}
G_{u,v}(l)
=
\begin{cases}
G_{\mathrm{main}},
& u\in\mathcal C_m(l),\ \psi_{u,v}(l)\le \Theta/2,\\[3pt]
G_{\mathrm{side}},
& u\in\mathcal C_m(l),\ \psi_{u,v}(l)>\Theta/2,\\[3pt]
0,
& u\notin\mathcal C_m(l),
\end{cases}
\label{eq:antenna_gain_comm}
\end{equation}
where $G_{\mathrm{main}}$ and $G_{\mathrm{side}}$ denote the main-lobe and side-lobe gains, respectively.

Let $P_u(l)$ denote the total ISAC transmit power of UAV $u$ for downlink communication and aerial-target sensing. Under line-of-sight (LoS) and free-space propagation, the communication channel power gain is $h_{u,v}^{\mathrm c}(l)=G_{u,v}(l)(\lambda/(4\pi d_{u,v}(l)))^2$, where $d_{u,v}(l)=\|\mathbf r_u(l)-\mathbf q_v(l)\|$ and $\lambda=c/f_c$, with $c$ and $f_c$ denoting the speed of light and carrier frequency, respectively. UAVs reuse the system bandwidth $B$ both within and across regions. Each UAV uniformly distributes its transmit power over $B$ and equally divides the bandwidth among its $k_u(l)$ served vehicles using orthogonal frequency resources. When UAV $u\in\mathcal U_m(l)$ serves vehicle $v\in\mathcal V_m(l)$, the complex-baseband signal received by vehicle $v$ over its allocated bandwidth $B/k_u(l)$ is
\begin{equation}
\begin{aligned}
y_v(t;l)
={}&\sqrt{\frac{P_u(l)}{k_u(l)}}\,
g_{u,v}^{\mathrm c}(l)s_{u,v}(t;l)\\
&\hspace{-2em}{}+
\sum_{j\in\mathcal C_m(l)\setminus\{u\}}
\sqrt{\frac{P_j(l)}{k_u(l)}}\,
g_{j,v}^{\mathrm c}(l)\widetilde{s}_{j,v}(t;l)
+n_v(t;l),
\end{aligned}
\label{eq:received_signal_comm}
\end{equation}
where $t$ denotes time within slot $l$, and $g_{j,v}^{\mathrm c}(l)$ is the complex channel coefficient with $|g_{j,v}^{\mathrm c}(l)|^2=h_{j,v}^{\mathrm c}(l)$. The zero-mean, mutually uncorrelated, unit-average-power waveforms $s_{u,v}$ and $\widetilde{s}_{j,v}$ represent the desired transmission and UAV $j$'s aggregate in-band interference, respectively. The noise is independent of the transmitted signals and has power $\mathbb E[|n_v(t;l)|^2]=N_0B/k_u(l)$, where $N_0$ is the additive white Gaussian noise (AWGN) power spectral density. The communication signal-to-interference-plus-noise ratio (SINR) is
\begin{equation}
\gamma_{u,v}^{\mathrm c}(l)
=
\frac{
\frac{P_u(l)}{k_u(l)}
h_{u,v}^{\mathrm c}(l)
}{
N_0\frac{B}{k_u(l)}
+
\sum\limits_{j\in\mathcal C_m(l)\setminus\{u\}}
\frac{P_j(l)}{k_u(l)}
h_{j,v}^{\mathrm c}(l)
},
\label{eq:sinr_comm}
\end{equation}
The achievable rate is $R_{u,v}^{\mathrm c}(l)=\frac{B}{k_u(l)}\log_2\!\left(1+\gamma_{u,v}^{\mathrm c}(l)\right)$, with minimum requirement $R_{\mathrm{th}}>0$. Vehicles are processed in descending order of their maximum SINR over candidate UAVs. Each vehicle is provisionally assigned to the UAV offering the highest projected rate after including it in bandwidth sharing. Each UAV then ranks its associated vehicles by descending SINR and serves the largest prefix whose rates all meet $R_{\mathrm{th}}$ under equal bandwidth sharing. The remaining vehicles are unserved in that slot.

Let $\chi_v(l)\in\{0,1\}$ denote the satisfaction indicator, where $\chi_v(l)=1$ if vehicle $v$'s achievable rate satisfies the minimum rate requirement $R_{\mathrm{th}}$, and $\chi_v(l)=0$ otherwise. Similar to \cite{bui2022robust}, the communication QoS of region $m$ is defined as:
\begin{equation}
Q_m(l)
=
\frac{1}{|\mathcal V_m(l)|}
\sum_{v\in\mathcal V_m(l)}
\chi_v(l),
\label{eq:communication_qos}
\end{equation}
where $Q_m(l)=0$ when $|\mathcal V_m(l)|=0$. In addition to regional QoS, slow-timescale cross-region UAV reallocation considers the overall service level and service balance across regions. The global communication QoS is defined as
\begin{equation}
Q(l)
=
\frac{
\sum_{m\in\mathcal M}
\sum_{v\in\mathcal V_m(l)}
\chi_v(l)
}{
\sum_{m\in\mathcal M}
|\mathcal V_m(l)|
},
\label{eq:global_communication_qos}
\end{equation}
where $Q(l)=0$ when $\sum_{m\in\mathcal M}|\mathcal V_m(l)|=0$. To distinguish the overall service level from its distribution across regions, regional service balance is measured using Jain's fairness index:
\begin{equation}
J(l)
=
\frac{
\left(\sum_{m\in\mathcal M}Q_m(l)\right)^2
}{
M\sum_{m\in\mathcal M}Q_m^2(l)+\varepsilon_J
},
\label{eq:qos_fairness}
\end{equation}
where $\varepsilon_J>0$ ensures numerical stability.

\subsection{Cross-Region Cooperative Sensing Utility}
When the target's horizontal projection lies in region $m$ at slot $l$, UAVs in $\mathcal C_m(l)$ perform distributed multiple-input multiple-output (MIMO) sensing using mutually orthogonal waveforms and coherently process the separated transmit-receive channels following~\cite{Godrich2009Phase}.

Under the synchronization assumptions in Section~II, region $m$ serves as the reference. Let $\epsilon_i(l)$ denote the residual phase offset of region $i$ relative to region $m$ at slot $l$, with $\epsilon_m(l)=0$ by definition. For each neighboring region $i\in\mathcal N_{\bar m}$, $\epsilon_i(l)$ is modeled as a zero-mean Gaussian random variable. Building on the synchronization-error decomposition in~\cite{Wang2022DistributedSAR}, we model the residual inter-region phase offset as
\begin{equation}
\resizebox{0.89\linewidth}{!}{$
\displaystyle
\epsilon_i(l)
=2\pi f_c\delta\tau_i(l)+2\pi T_{\mathrm c}\delta f_i(l)+\epsilon_{\mathrm{pn},i}(l)+\epsilon_{\mathrm{cal},i}(l),
$}
\label{eq:residual_phase_decomposition}
\end{equation}
where $\delta\tau_i(l)$ and $\delta f_i(l)$ denote the residual timing and carrier-frequency offsets relative to region $m$, respectively, and $\epsilon_{\mathrm{pn},i}(l)$ denotes the relative oscillator phase noise. The term $\epsilon_{\mathrm{cal},i}(l)$ captures the remaining hardware and initial-phase calibration residual, excluding the errors already represented by the other three terms. The residual carrier-frequency offset is assumed constant over the coherent processing interval $T_{\mathrm c}$, and its phase accumulation is evaluated at the end of that interval. Assuming that the four residual components are zero-mean and mutually independent, their variances add to give
\begin{equation}
\resizebox{0.89\linewidth}{!}{$
\displaystyle
\sigma_{\epsilon,i}^{2}(l)
=
(2\pi f_c)^2\sigma_{\tau,i}^{2}(l)
+
(2\pi T_{\mathrm c})^2\sigma_{f,i}^{2}(l)
+
\sigma_{\mathrm{pn},i}^{2}(l)
+
\sigma_{\mathrm{cal},i}^{2}(l),
$}
\label{eq:expanded_phase_variance}
\end{equation}
where $\sigma_{\tau,i}^{2}(l)$, $\sigma_{f,i}^{2}(l)$, $\sigma_{\mathrm{pn},i}^{2}(l)$, and $\sigma_{\mathrm{cal},i}^{2}(l)$ are the variances of $\delta\tau_i(l)$, $\delta f_i(l)$, $\epsilon_{\mathrm{pn},i}(l)$, and $\epsilon_{\mathrm{cal},i}(l)$, respectively. Inter-region handshaking exchanges timing, carrier-frequency, and phase-reference information between regions, while the target-dependent propagation phase is retained for localization.

\begin{figure}[!t]
\centering
\includegraphics[width=\columnwidth]{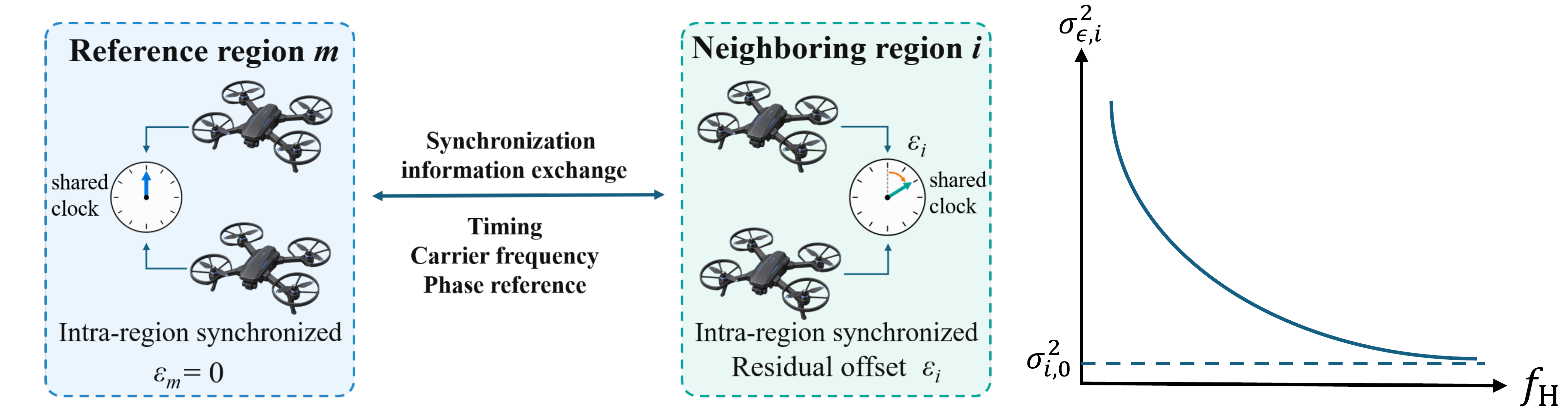}
\caption{Conceptual illustration of inter-region UAV handshaking for cooperative sensing.}
\label{fig:handshaking_mechanism}
\end{figure}

Fig.~\ref{fig:handshaking_mechanism} illustrates the exchange of synchronization-reference information between UAVs in different regions. Inter-region handshaking periodically updates the synchronization state between regions. Between two consecutive updates, random clock and oscillator drift accumulates with the update interval, while the phase error caused by a residual carrier-frequency offset grows linearly with that interval. Since the interval between consecutive updates is inversely proportional to $f_{\mathrm H}(l)$, the residual inter-region phase-error variance is modeled as
\begin{equation}
\sigma_{\epsilon,i}^{2}(l)
=
\sigma_{i,0}^{2}
+ \frac{\kappa_{i,1}}{f_{\mathrm H}(l)}
+ \frac{\kappa_{i,2}}{f_{\mathrm H}^{2}(l)},
\quad f_{\mathrm H}(l)>0,\quad
i\in\mathcal N_{\bar m},
\label{eq:handshake_phase_variance}
\end{equation}
where $f_{\mathrm H}(l)$ is the common handshaking frequency among the participating regions, and $\sigma_{i,0}^{2}$ denotes the handshaking-frequency-independent residual uncertainty immediately after an update. The coefficients $\kappa_{i,1}$ and $\kappa_{i,2}$ are positive. The term $\kappa_{i,1}/f_{\mathrm H}(l)$ captures the accumulated random drift, whereas $\kappa_{i,2}/f_{\mathrm H}^{2}(l)$ captures the phase uncertainty caused by residual carrier-frequency offset. Thus, a lower $f_{\mathrm H}(l)$ represents weaker inter-region synchronization, while $f_{\mathrm H}(l)=0$ denotes no inter-region handshaking.

Let $\boldsymbol{\epsilon}(l)$ collect the neighboring regions' residual phase offsets relative to region $m$. Assuming independent offsets, their covariance is $\boldsymbol{\Sigma}_{\epsilon}(l)=\operatorname{diag}\{\sigma_{\epsilon,i}^{2}(l)\}_{i\in\mathcal N_{\bar m}}$. Following~\cite{Godrich2009Phase}, $\mathbf J_{\mathbf p\mathbf p}(l)$, $\mathbf J_{\mathbf p\boldsymbol\epsilon}(l)$, and $\mathbf J_{\boldsymbol\epsilon\boldsymbol\epsilon}(l)$ are the sensing-data Fisher information matrix (FIM) blocks for $\mathbf p(l)$ and $\boldsymbol{\epsilon}(l)$ after jointly eliminating the shared unknown target reflectivity across all included paths. With $\mathbf p(l)$ unknown but fixed within each slot and $\boldsymbol{\epsilon}(l)$ random, the reduced hybrid FIM is
\begin{equation}
\mathbf J_{\mathrm H}(l)
=
\begin{bmatrix}
\mathbf J_{\mathbf p\mathbf p}(l)
&
\mathbf J_{\mathbf p\boldsymbol\epsilon}(l)
\\
\mathbf J_{\mathbf p\boldsymbol\epsilon}^{T}(l)
&
\mathbf J_{\boldsymbol\epsilon\boldsymbol\epsilon}(l)
+
\boldsymbol{\Sigma}_{\epsilon}^{-1}(l)
\end{bmatrix}.
\label{eq:hybrid_fim}
\end{equation}
The prior precision $\boldsymbol{\Sigma}_{\epsilon}^{-1}(l)$ is set to zero when $f_{\mathrm H}(l)=0$, indicating no handshaking-derived prior information. For $f_{\mathrm H}(l)>0$, eliminating $\boldsymbol{\epsilon}(l)$ via the Schur complement gives the equivalent position FIM:
\begin{equation}
\mathbf J_{\mathbf p}(l)
=
\mathbf J_{\mathbf p\mathbf p}(l)
-
\mathbf J_{\mathbf p\boldsymbol\epsilon}(l)
\left(
\mathbf J_{\boldsymbol\epsilon\boldsymbol\epsilon}(l)
+
\boldsymbol{\Sigma}_{\epsilon}^{-1}(l)
\right)^{-1}
\mathbf J_{\mathbf p\boldsymbol\epsilon}^{T}(l).
\label{eq:equivalent_position_fim}
\end{equation}
Sensing performance is measured by the localization CRB:
\begin{equation}
\Phi_{\mathrm{CRB}}(l)
=
\operatorname{Tr}
\left[
\mathbf J_{\mathbf p}^{-1}(l)
\right].
\label{eq:sensing_crb}
\end{equation}
A smaller CRB yields a higher cooperative sensing utility:
\begin{equation}
U_{\mathrm s}(l)
=
\frac{
\log_{10}\!\left(
\Phi_{\max}/\Phi_{\mathrm{CRB}}(l)
\right)
}{
\log_{10}\!\left(
\Phi_{\max}/\Phi_{\min}
\right)
},
\label{eq:sensing_utility}
\end{equation}
where $0<\Phi_{\min}<\Phi_{\max}$ are normalization bounds satisfying $\Phi_{\min}\le\Phi_{\mathrm{CRB}}(l)\le\Phi_{\max}$.

\begin{proposition}[Establishment of Optimal Handshaking Frequency]
\label{prop:handshaking_frequency}
For a fixed sensing configuration with handshaking energy $E_{\mathrm H}(f_{\mathrm H})=c_{\mathrm H}f_{\mathrm H}$ and relative handshaking energy weight $\eta$, any interior maximizer $f_{\mathrm H}^{\star}$ of the overall system utility satisfies $\left.(\partial U_{\mathrm s}/\partial f_{\mathrm H})\right|_{f_{\mathrm H}=f_{\mathrm H}^{\star}}=\eta c_{\mathrm H}$.
\end{proposition}

\begin{IEEEproof}
Omit the slot index $l$ for brevity. For the fixed sensing configuration, the sensing-data FIM blocks are independent of $f_{\mathrm H}$. From \eqref{eq:handshake_phase_variance}, the residual phase-error variance satisfies $\partial\sigma_{\epsilon,i}^{2}/\partial f_{\mathrm H}=-\kappa_{i,1}/f_{\mathrm H}^{2}-2\kappa_{i,2}/f_{\mathrm H}^{3}<0$. Define $\mathbf K(f_{\mathrm H})=\mathbf J_{\boldsymbol\epsilon\boldsymbol\epsilon}+\boldsymbol\Sigma_{\epsilon}^{-1}(f_{\mathrm H})$. For $f_2>f_1>0$, we have $\mathbf K(f_2)\succ\mathbf K(f_1)\succ\mathbf0$. Hence, \eqref{eq:equivalent_position_fim} gives
\begin{equation}
\resizebox{0.89\linewidth}{!}{$
\begin{aligned}
&\mathbf J_{\mathbf p}(f_2)-\mathbf J_{\mathbf p}(f_1)=\mathbf J_{\mathbf p\boldsymbol\epsilon}
\bigl[\mathbf K^{-1}(f_1)-\mathbf K^{-1}(f_2)\bigr]
\mathbf J_{\mathbf p\boldsymbol\epsilon}^{T}\succeq\mathbf0.
\end{aligned}
$}
\label{eq:position_fim_order}
\end{equation}
This increment is nonzero because $\mathbf J_{\mathbf p\boldsymbol\epsilon}\ne\mathbf0$. For positive-definite position FIMs, the inverse difference $\mathbf J_{\mathbf p}^{-1}(f_1)-\mathbf J_{\mathbf p}^{-1}(f_2)$ is also positive semidefinite and nonzero, yielding $\Phi_{\mathrm{CRB}}(f_2)<\Phi_{\mathrm{CRB}}(f_1)$ and $U_{\mathrm s}(f_2)>U_{\mathrm s}(f_1)$.

Meanwhile, the aggregate normalized handshaking energy of all participating UAVs is $E_{\mathrm H}(f_{\mathrm H})=c_{\mathrm H}f_{\mathrm H}$, where $c_{\mathrm H}>0$ is constant for the fixed configuration and system energy normalization. The detailed handshaking energy model is presented in the following subsection. Let $U(f_{\mathrm H})=U_{\mathrm s}(f_{\mathrm H})-\eta E_{\mathrm H}(f_{\mathrm H})$ denote the frequency-dependent part of the per-slot system utility, where $\eta\ge0$ is the relative handshaking energy weight. Then $U(f_2)-U(f_1)=U_{\mathrm s}(f_2)-U_{\mathrm s}(f_1)-\eta c_{\mathrm H}(f_2-f_1)$. Therefore, when
\begin{equation}
\eta c_{\mathrm H}>
\frac{U_{\mathrm s}(f_2)-U_{\mathrm s}(f_1)}{f_2-f_1},
\label{eq:handshaking_energy_threshold}
\end{equation}
it has $U(f_2)<U(f_1)$ despite the CRB reduction. At any interior maximizer $f_{\mathrm H}^{\star}$, the first-order necessary condition $U'(f_{\mathrm H}^{\star})=0$ yields $\left.(\partial U_{\mathrm s}/\partial f_{\mathrm H})\right|_{f_{\mathrm H}=f_{\mathrm H}^{\star}}=\eta c_{\mathrm H}$.
\end{IEEEproof}

Proposition~\ref{prop:handshaking_frequency} shows that more frequent handshaking improves sensing accuracy at the cost of higher energy consumption. This tradeoff motivates adapting the handshaking frequency to the current sensing configuration and the relative energy cost.

To quantify the sensing contribution of each region in $\mathcal N_m$, let $\Phi_{\mathrm{CRB}}^{(-i)}(l)$ denote the position CRB recomputed after removing all UAVs in region $i$ from both the transmitting and receiving sets, while retaining all other participating regions. The regional marginal CRB utility of region $i$ is defined as
\begin{equation}
U_{\mathrm s}^{i}(l)
=
U_{\mathrm s}(l)
-
\frac{
\log_{10}\!\left(
\Phi_{\max}/\Phi_{\mathrm{CRB}}^{(-i)}(l)
\right)
}{
\log_{10}\!\left(
\Phi_{\max}/\Phi_{\min}
\right)
},
\quad
i\in\mathcal N_m.
\label{eq:marginal_crb_utility}
\end{equation}
A larger $U_{\mathrm s}^{i}(l)$ indicates that excluding region $i$ causes a larger degradation in localization accuracy and, hence, that region $i$ provides a greater marginal contribution to cross-region cooperative sensing.

\subsection{UAV-Swarm Energy Utility}
The communication and sensing benefits described above incur energy costs. UAV operation consumes energy for propulsion, unified ISAC transmission, and inter-region handshaking. The last component explicitly accounts for the cost of reducing the synchronization uncertainty.

For UAV $u$, let $\mathbf v_u(l)=[v_{u,x}(l),v_{u,y}(l),v_{u,z}(l)]^T$ denote its three-dimensional velocity, where $v_{u,x}(l)$ and $v_{u,y}(l)$ are the horizontal velocity components, and $v_{u,z}(l)$ is the vertical velocity. Following the generic three-dimensional multirotor power model in~\cite{Gong2023Power}, the propulsion power under steady flight is modeled as $P_u^{\mathrm f}(l)=P_u^{\mathrm h}\!\left(v_{u,x}(l),v_{u,y}(l)\right)+\Delta P_u^{\mathrm v}\!\left(v_{u,z}(l)\right)$. Here, $P_u^{\mathrm h}(\cdot)$ accounts for the hovering and horizontal-motion power components, while $\Delta P_u^{\mathrm v}(\cdot)$ represents the power adjustment caused by vertical motion. The latter follows different expressions for ascent and descent according to the sign of $v_{u,z}(l)$, with $\Delta P_u^{\mathrm v}(0)=0$. Accordingly, the propulsion energy consumed by UAV $u$ during slot $l$ is $E_u^{\mathrm f}(l)=P_u^{\mathrm f}(l)\Delta t$. Since each UAV transmits a unified ISAC signal for both communication and sensing, its transmit power is counted only once, and the corresponding transmission energy is $E_u^{\mathrm{tx}}(l)=P_u(l)\Delta t$.

Inter-region handshaking exchanges timing, frequency, and phase-reference information among UAVs in the participating regions. Assuming that the aerial target is located in region $i$ at slot $l$, the handshaking energy of UAV $u$ is modeled as
\begin{equation}
E_u^{\mathrm H}(l)
=
\mathbf 1_{\{u\in\mathcal C_i(l)\}}
\zeta_{\mathrm H}f_{\mathrm H}(l)\Delta t,
\label{eq:handshake_energy}
\end{equation}
where $\zeta_{\mathrm H}$ denotes the energy consumed by each participating UAV in one handshaking update. Therefore, the normalized energy consumption of region $m$ is defined as
\begin{equation}
E_m(l)
=
\frac{
\sum_{u\in\mathcal U_m(l)}
\left[
E_u^{\mathrm f}(l)+
E_u^{\mathrm{tx}}(l)+
E_u^{\mathrm H}(l)
\right]
}{
E_m^{\max}
},
\label{eq:regional_energy_utility}
\end{equation}
where $E_m^{\max}$ is a reference energy used for normalization.

\section{Problem Formulation and Two-Timescale Decomposition}
\label{sec:problem_formulation}
Using the utilities derived in Section~III, this section first formulates the joint optimization problem for cross-region cooperation. We then present two coupled subproblems for slow-timescale UAV reallocation and fast-timescale UAV control.

\subsection{Joint Optimization of Cross-Region UAV Cooperation}
\label{subsec:cross_region_cooperation}

For reallocation round $n$, let $\mathbf F_n=[f_{ij,n}]_{M\times M}$ denote the binary dispatch matrix, where $f_{ij,n}=1$ means that region $i$ sends one available UAV to neighboring region $j$. Diagonal and non-neighboring entries are zero, and a zero row indicates no dispatch. Let $\mathbf u_n=[u_{i,n}]_{i\in\mathcal M}$ identify the selected UAVs, with $u_{i,n}=0$ for a non-dispatching region. Together, $(\mathbf F_n,\mathbf u_n)$ determine the allocation under which UAVs adjust their mobility, transmit power, and handshaking preferences.

Let $\mathbf J_{\mathbf p}^{\mathrm{loc}}(l)\succ\mathbf0$ denote the equivalent position FIM obtained using only UAVs in the target's region, with corresponding localization CRB $\Phi_{\mathrm{CRB}}^{\mathrm{loc}}(l)=\operatorname{Tr}[(\mathbf J_{\mathbf p}^{\mathrm{loc}}(l))^{-1}]$. The benefit of neighboring-region observations is characterized below with the local sensing configuration held fixed.

\begin{proposition}[Sensing Error Reduction via Cross-Region Cooperation]
\label{prop:cross_region_crb}
For a neighboring UAV $j$ with effective monostatic ranging information $a_j>0$ and unit UAV-to-target direction $\mathbf e_j$, cooperation with this UAV at a fixed handshaking frequency guarantees a CRB reduction of at least $\frac{a_j\mathbf e_j^T(\mathbf J_{\mathbf p}^{\mathrm{loc}})^{-2}\mathbf e_j}{1+a_j\mathbf e_j^T(\mathbf J_{\mathbf p}^{\mathrm{loc}})^{-1}\mathbf e_j}$ relative to local-only sensing.
\end{proposition}

\begin{IEEEproof}
Omit the slot index $l$. The monostatic echo provides delay information $\mathbf D_j=a_j\mathbf e_j\mathbf e_j^T$ that remains after eliminating the unknown reflectivity and regional phase offsets. In a common full parameter space, assign the unchanged phase priors to the local term. Separately eliminating the nuisance parameters for the retained observations and the independent echo relaxes their shared nuisance parameters, giving $\mathbf J_{\mathbf p}\succeq\mathbf J_{\mathbf p}^{\mathrm{loc}}+\mathbf D_j$. Matrix inversion reverses this ordering, and the Sherman-Morrison formula yields
\begin{equation}
\begin{aligned}
\Phi_{\mathrm{CRB}}
&\le\operatorname{Tr}\!\left[
(\mathbf J_{\mathbf p}^{\mathrm{loc}}+a_j\mathbf e_j\mathbf e_j^T)^{-1}
\right]\\
&=\Phi_{\mathrm{CRB}}^{\mathrm{loc}}
-\frac{
a_j\mathbf e_j^T(\mathbf J_{\mathbf p}^{\mathrm{loc}})^{-2}\mathbf e_j
}{
1+a_j\mathbf e_j^T(\mathbf J_{\mathbf p}^{\mathrm{loc}})^{-1}\mathbf e_j
}.
\end{aligned}
\end{equation}
Rearranging gives the stated lower bound on the CRB reduction, which is strictly positive since $a_j>0$, $\|\mathbf e_j\|=1$, and $\mathbf J_{\mathbf p}^{\mathrm{loc}}\succ\mathbf0$.
\end{IEEEproof}

Proposition~\ref{prop:cross_region_crb} highlights the value of neighboring observations that provide strong ranging information along directions poorly resolved by the local swarm. Together with the sensing-energy tradeoff established in Proposition~\ref{prop:handshaking_frequency}, this result motivates jointly adapting UAV sensing geometry and handshaking frequency while accounting for communication requirements.

UAV handshaking preferences $f_{u,\mathrm H}(l)$ are aggregated into the common frequency $f_{\mathrm H}(l)$ as described in Section~V. Let $E_{\mathrm{all}}(l)$ denote the normalized total energy of cross-region dispatch and intra-region control. The corresponding energy costs are defined in~\eqref{eq:total_dispatch_energy} and~\eqref{eq:regional_energy_utility}, respectively. With nonnegative weights $\omega_1$, $\omega_2$, $\omega_3$, and $\omega_4$, the overall problem is
\begin{subequations}
\label{problem_cross_region_cooperation}
\begin{IEEEeqnarray}{-l}
\parbox[t]{\columnwidth}{\raggedright\bfseries\settowidth{\hangindent}{P0:\ }\hangafter=1 P0:\ Joint Inter-Region Reallocation and Intra-Region Control Optimization\par}
\nonumber
\\[-0.3em]
\resizebox{\linewidth}{!}{$\displaystyle
\begin{aligned}
\underset{
\substack{
\{\mathbf F_n,\mathbf u_n\},\\
\{\mathbf v_u(l),P_u(l),f_{u,\mathrm H}(l)\}
}
}{\max}
\quad
&
\sum_{l\in\mathcal T_{\mathrm f}}
\left[
\omega_1 Q(l)
+
\omega_2 J(l)
+
\omega_3 U_{\mathrm s}(l)
-
\omega_4 E_{\mathrm{all}}(l)
\right]
\end{aligned}
$}
\nonumber
\\
\hspace{2.4em} \mathrm{s.t.}\quad \quad \quad
\left\{
\mathbf F_n,
\mathbf u_n,
\mathbf v_u(l),
P_u(l),
f_{u,\mathrm H}(l)
\right\}
\in\mathcal X,
\label{eq:p0_constraints}
\end{IEEEeqnarray}
\end{subequations}
where constraint~\eqref{eq:p0_constraints} requires the reallocation and control variables to satisfy their corresponding feasibility conditions. Here, $\mathcal X$ denotes their joint feasible set.

Problem \textbf{P0} involves mixed discrete and continuous decisions, long-term performance optimization, and nonconvex objectives and constraints. The dispatch matrix $\mathbf F_n$ and UAV selection vector $\mathbf u_n$ are determined once per reallocation round, whereas UAV velocities $\mathbf v_u(l)$, transmit powers $P_u(l)$, and handshaking preferences $f_{u,\mathrm H}(l)$ are updated at each slot. To coordinate these coupled decisions at their respective timescales, we decompose \textbf{P0} into a slow-timescale UAV reallocation problem \textbf{P1} and a fast-timescale intra-region control problem \textbf{P2}. In \textbf{P1}, UAV reallocation prioritizes regional QoS and fairness subject to a sensing-loss tolerance, while \textbf{P2} jointly optimizes communication and sensing performance under the resulting allocation.

\subsection{Slow-Timescale Cross-Region Reallocation Problem}
\label{subsec:slow_timescale_scheduling_problem}

Let $\mathcal U_{i,n}$ be the UAVs available in region $i$ at $l=(n-1)K+1$. For candidate $u$, let $\Phi_{\mathrm{CRB}}^{(-u)}(l)$ be the CRB after removing its transmit-receive paths while fixing all remaining positions, powers, and synchronization statistics. Define $L_{u,n}^{\mathrm s}=\log_{10}(\Phi_{\mathrm{CRB}}^{(-u)}(l)/\Phi_{\mathrm{CRB}}(l))$ and the sensing-safe set $\mathcal U_{i,n}^{\mathrm{safe}}=\{u\in\mathcal U_{i,n}:L_{u,n}^{\mathrm s}\le\Gamma_{\mathrm s}\}$, where $\Gamma_{\mathrm s}$ is the tolerated logarithmic CRB increase. Candidates causing unidentifiability are excluded. The signed communication-removal loss is $L_{i,u,n}^{\mathrm c}=Q_i(l)-Q_i^{(-u)}(l)$, where $Q_i^{(-u)}(l)$ is the source QoS without UAV $u$. Let $E_{u,i,j,n}^{\mathrm d}$ denote propulsion energy accumulated along UAV $u$'s transit from region $i$ to the designated arrival point in region $j$, normalized by $E_{\mathrm{ref}}^{\mathrm d}$. The round-level dispatch energy is
\begin{equation}
\resizebox{0.89\linewidth}{!}{$\displaystyle
E_n^{\mathrm d}
=\frac{1}{U_{\mathrm{dis}}^{\max}}
\sum_{i\in\mathcal M}\sum_{j\in\mathcal N_{\bar i}}
f_{ij,n}E_{u_{i,n},i,j,n}^{\mathrm d}, U_{\mathrm{dis}}^{\max}=\lfloor M/2\rfloor,
$}
\label{eq:total_dispatch_energy}
\end{equation}
where $U_{\mathrm{dis}}^{\max}=\lfloor M/2\rfloor$ follows from regional role exclusivity. Round-level QoS and fairness are averaged over the last $W$ slots before the next reallocation decision, with $1\le W\le K$:%
\begin{subequations}
\label{eq:slow_timescale_performance}
\begin{align}
Q_n &= \frac{1}{W}\sum_{l=n K-W+1}^{n K} Q(l),
\label{eq:slow_timescale_qos}\\
J_n &= \frac{1}{W}\sum_{l=n K-W+1}^{n K} J(l).
\label{eq:slow_timescale_fairness}
\end{align}
\end{subequations}

The coefficients $a_1$, $a_2$, and $a_3$ are nonnegative weighting coefficients for global communication QoS, regional service fairness, and dispatch energy, respectively. The slow-timescale cross-region UAV reallocation problem is formulated as
\begin{subequations}
\label{problem_slow_scheduling}
\begin{IEEEeqnarray}{-l}
\parbox[t]{\columnwidth}{\raggedright\bfseries\settowidth{\hangindent}{P1:\ }\hangafter=1 P1:\ Inter-Region Reallocation for Communication and Energy Optimization\par}
\nonumber
\\[-0.3em]
\begin{aligned}
\underset{\{\mathbf F_n,\mathbf u_n\}}{\max}
\quad
&
\sum_{n\in\mathcal T_{\mathrm s}}
\left[
a_1 Q_n
+
a_2 J_n
-
a_3E_{n}^{\mathrm d}
\right]
\end{aligned}
\nonumber
\\
\enspace\mathrm{s.t.}\quad
f_{ij,n}\in\{0,1\},
\quad
\forall i\in\mathcal M,\;
j\in\mathcal N_{\bar i},
\label{eq:slow_binary_action}
\\
\phantom{\enspace\mathrm{s.t.}\quad}
\sum_{j\in\mathcal N_{\bar i}}
f_{ij,n}
+
\sum_{k:\,i\in\mathcal N_{\bar k}}
f_{ki,n}
\le 1,
\quad
\forall i\in\mathcal M,
\label{eq:slow_role_exclusivity}
\\
\phantom{\enspace\mathrm{s.t.}\quad}
u_{i,n}\in\mathcal U_{i,n}^{\mathrm{safe}},
\quad \forall i:\sum_{j\in\mathcal N_{\bar i}}f_{ij,n}=1,
\label{eq:slow_sensing_safe}
\\
\phantom{\enspace\mathrm{s.t.}\quad}
\left|\mathcal U_{i,n}\right|
-
\sum_{j\in\mathcal N_{\bar i}}
f_{ij,n}
\ge 1,
\quad
\forall i\in\mathcal M,\;
n\in\mathcal T_{\mathrm s},
\label{eq:slow_uav_reserve}
\end{IEEEeqnarray}
\end{subequations}
where constraint \eqref{eq:slow_binary_action} restricts dispatch decisions between neighboring regions to binary values. Constraint \eqref{eq:slow_role_exclusivity} allows each region to dispatch at most one UAV or receive at most one UAV and prevents it from acting as both a source and a destination in the same reallocation round. Constraint \eqref{eq:slow_sensing_safe} requires the UAV selected for each dispatch to satisfy the sensing-removal tolerance. Constraint \eqref{eq:slow_uav_reserve} ensures that each source region retains at least one available UAV after dispatch.

\subsection{Fast-Timescale Intra-Region Control Problem}
\label{subsec:fast_timescale_control}

Given the UAV allocation determined by \textbf{P1}, each region controls UAV operation to improve local communication QoS while reducing the global CRB and energy consumption. Let $\alpha_m(l)\in\{0,1\}$ indicate whether region $m$ participates in the cross-region cooperative sensing task at slot $l$. Specifically, $\alpha_m(l)=1$ for the region containing the target and its participating neighboring regions, and $\alpha_m(l)=0$ otherwise. Let $\mathbf r_u^{xy}(l)$ denote the horizontal position of UAV $u$, and let $\Omega_m$ denote the horizontal area of region $m$. The coefficients $b_1$, $b_2$, and $b_3$ are nonnegative weighting coefficients for regional communication QoS, cooperative sensing utility, and regional energy consumption, respectively. Let $V_{\mathrm{xy}}^{\max}$ denote the maximum magnitude of each horizontal velocity component, and let $V_{\mathrm z}^{\max}$ denote the maximum vertical speed. Moreover, $P_{\max}$, $[f_{\mathrm H}^{\min},f_{\mathrm H}^{\max}]$, $[z_{\min},z_{\max}]$, and $d_{\min}$ denote the maximum transmit power, allowable UAV-level handshaking preference range, allowable altitude range, and minimum safety distance, respectively. The fast-timescale intra-region control problem of region $m$ is formulated as
\begin{subequations}
\label{problem_fast_control}
\begin{IEEEeqnarray}{-l}
\parbox[t]{\columnwidth}{\raggedright\bfseries\settowidth{\hangindent}{P2:\ }\hangafter=1 P2:\ Intra-Region Control for Communication, Sensing, and Energy Optimization\par}
\nonumber
\\[-0.3em]
\scalebox{0.83}{$\displaystyle
\begin{aligned}
\underset{
\{\mathbf v_u(l),P_u(l),f_{u,\mathrm H}(l)\}
}{\max}
\quad
&
\sum_{l\in\mathcal T_{\mathrm f}}
\left[
b_1Q_m(l)
+
b_2\alpha_m(l)U_{\mathrm s}(l)
-
b_3E_m(l)
\right]
\end{aligned}
$}
\nonumber
\\
\enspace\mathrm{s.t.}\quad
\left|v_{u,x}(l)\right|, \left|v_{u,y}(l)\right|
\le
V_{\mathrm{xy}}^{\max},
\quad
\forall u\in\mathcal U_m(l),
\label{eq:control_horizontal_velocity}
\\
\phantom{\enspace\mathrm{s.t.}\quad}
\left|v_{u,z}(l)\right|
\le
V_{\mathrm z}^{\max},
\quad
\forall u\in\mathcal U_m(l),
\label{eq:control_vertical_velocity}
\\
\phantom{\enspace\mathrm{s.t.}\quad}
0
\le
P_u(l)
\le
P_{\max},
\quad
\forall u\in\mathcal U_m(l),
\label{eq:control_power}
\\
\phantom{\enspace\mathrm{s.t.}\quad}
f_{\mathrm H}^{\min}
\le
f_{u,\mathrm H}(l)
\le
f_{\mathrm H}^{\max},
\forall u\in\mathcal U_m(l),
\label{eq:control_handshake}
\\
\phantom{\enspace\mathrm{s.t.}\quad}
z_{\min}
\le
z_u(l)
\le
z_{\max},
\quad
\forall u\in\mathcal U_m(l),
\label{eq:control_altitude}
\\
\phantom{\enspace\mathrm{s.t.}\quad}
\mathbf r_u^{xy}(l)
\in
\Omega_m,
\quad
\forall u\in\mathcal U_m(l),
\label{eq:control_region_boundary}
\\
\phantom{\enspace\mathrm{s.t.}\quad}
\scalebox{0.95}{$\displaystyle
\left\|
\mathbf r_u(l)-\mathbf r_j(l)
\right\|
\ge
d_{\min},
\quad
\forall u,j\in\mathcal U_m(l),\ u\neq j
$},
\label{eq:control_safety_distance}
\end{IEEEeqnarray}
\end{subequations}
where constraint \eqref{eq:control_horizontal_velocity} limits each horizontal velocity component of a UAV in region $m$. Constraint \eqref{eq:control_vertical_velocity} bounds its vertical velocity. Constraint \eqref{eq:control_power} limits its ISAC transmit power. Constraint \eqref{eq:control_handshake} specifies the feasible range of each UAV-level handshaking preference. Constraint \eqref{eq:control_altitude} imposes the allowable altitude range. Constraint \eqref{eq:control_region_boundary} keeps its horizontal position within the assigned region. Constraint \eqref{eq:control_safety_distance} enforces minimum intra-region UAV separation.

\section{Cross-Region Cooperation of UAV Swarms via Hierarchical MARL}
\label{sec:methodology}
In this section, we first explain the motivation for the H-MARL framework and introduce its stagewise training strategy, then describe graph-based independent proximal policy optimization (IPPO) for regional reallocation and multi-agent proximal policy optimization (MAPPO) for intra-region UAV control, and finally present policy optimization.

\subsection{Motivation for Hierarchical MARL}
\label{subsec:motivation_hmarl}
Although the two-timescale strategy separates regional reallocation from slot-level control, \textbf{P1} still involves combinatorial dispatch and UAV selection, while \textbf{P2} retains nonconvex objectives and constraints. Reallocation and mobility also affect future UAV resources and spatial configurations. We therefore adopt reinforcement learning to learn allocation and control policies, accounting for delayed effects through cumulative rewards. Regional and UAV decisions interact through resource redistribution, communication interference, and cooperative sensing, requiring coordination among agents. A centralized single-agent policy would face a joint action space that grows with the number of regions and UAVs. MARL instead assigns local decisions to individual agents and guides coordination through shared team or regional rewards. Regional and UAV agents are organized hierarchically to implement the two-timescale strategy for \textbf{P1} and \textbf{P2}.

The slow-timescale transition and reward mappings depend on the fast-timescale control policy. We adopt stagewise training~\cite{Florensa2017SNN}: the fast-timescale MAPPO policy is pretrained and then held fixed during slow-timescale graph-based IPPO training. A paired All-Stay reference starts from the same initial state and retains the initial regional UAV allocation while applying the same fast-timescale policy. Let $\mathbf s_n^{\mathrm s}$ denote the full slow-timescale state, including the physical system state, the paired reference state, and the round index. Let $\mathbf a_n^{\mathrm s}=(\mathbf F_n,\mathbf u_n)$ denote the complete joint reallocation action. The local graph observations need not contain this full state. Let $e$ index the updates of the slow-timescale policy. For reallocation round $n$, let $p_{\mathrm s}^{[e]}(\mathbf s_{n+1}^{\mathrm s}|\mathbf s_n^{\mathrm s},\mathbf a_n^{\mathrm s})$ denote the induced transition probability from the current system state $\mathbf s_n^{\mathrm s}$ to the next system state
$\mathbf s_{n+1}^{\mathrm s}$, and let $\bar r_{i,n}^{\mathrm s,[e]}(\mathbf s_n^{\mathrm s},\mathbf a_n^{\mathrm s})$ denote the corresponding expected reward of reallocation agent $i$.

\begin{proposition}[Slow-Timescale Stationarity]
\label{prop:stagewise_stationarity}
If the environment transition law, reward calculation, and fast-timescale behavior policy remain fixed during slow-timescale training, the induced slow-timescale transition probabilities and expected rewards are independent of the slow-timescale policy update index $e$, i.e.,
\begin{subequations}
\label{eq:stagewise_stationarity}
\begin{align}
p_{\mathrm s}^{[e+1]}(\mathbf s_{n+1}^{\mathrm s}|
\mathbf s_n^{\mathrm s},\mathbf a_n^{\mathrm s})
&=
p_{\mathrm s}^{[e]}(\mathbf s_{n+1}^{\mathrm s}|
\mathbf s_n^{\mathrm s},\mathbf a_n^{\mathrm s}),
\label{eq:stagewise_transition_stationarity}
\\
\bar r_{i,n}^{\mathrm s,[e+1]}
(\mathbf s_n^{\mathrm s},\mathbf a_n^{\mathrm s})
&=
\bar r_{i,n}^{\mathrm s,[e]}
(\mathbf s_n^{\mathrm s},\mathbf a_n^{\mathrm s}).
\label{eq:stagewise_reward_stationarity}
\end{align}
\end{subequations}
\end{proposition}

\begin{IEEEproof}
Let $\boldsymbol{\zeta}_n$ collect the policy and paired All-Stay fast-timescale state-action trajectories used to evaluate reallocation round $n$. At slow-timescale policy update $e$, its conditional distribution is denoted by $p^{[e]}(\boldsymbol{\zeta}_n|\mathbf s_n^{\mathrm s},\mathbf a_n^{\mathrm s})$. Since the environment transition law and the fast-timescale behavior policy remain fixed,
\begin{equation}
p^{[e+1]}(\boldsymbol{\zeta}_n|
\mathbf s_n^{\mathrm s},\mathbf a_n^{\mathrm s})
=
p^{[e]}(\boldsymbol{\zeta}_n|
\mathbf s_n^{\mathrm s},\mathbf a_n^{\mathrm s}).
\label{eq:fixed_fast_trajectory_distribution}
\end{equation}

The induced slow-timescale transition probability and expected reward are given by
\begin{subequations}
\label{eq:induced_slow_transition_reward}
\begin{equation}
\scalebox{0.85}{$\displaystyle
p_{\mathrm s}^{[e]}(\mathbf s_{n+1}^{\mathrm s}|
\mathbf s_n^{\mathrm s},\mathbf a_n^{\mathrm s})
=
\int
p(\mathbf s_{n+1}^{\mathrm s}|
\boldsymbol{\zeta}_n,\mathbf s_n^{\mathrm s},\mathbf a_n^{\mathrm s})
p^{[e]}(\boldsymbol{\zeta}_n|
\mathbf s_n^{\mathrm s},\mathbf a_n^{\mathrm s})
\,\mathrm d\boldsymbol{\zeta}_n,
$}
\label{eq:induced_slow_transition}
\end{equation}
\begin{equation}
\scalebox{0.97}{$\displaystyle
\bar r_{i,n}^{\mathrm s,[e]}
(\mathbf s_n^{\mathrm s},\mathbf a_n^{\mathrm s})
=
\int
r_{i,n}^{\mathrm s}
(\mathbf s_n^{\mathrm s},\mathbf a_n^{\mathrm s},\boldsymbol{\zeta}_n)
p^{[e]}(\boldsymbol{\zeta}_n|
\mathbf s_n^{\mathrm s},\mathbf a_n^{\mathrm s})
\,\mathrm d\boldsymbol{\zeta}_n.
$}
\label{eq:induced_slow_reward}
\end{equation}
\end{subequations}
Here, $p(\mathbf s_{n+1}^{\mathrm s}|\boldsymbol{\zeta}_n,\mathbf s_n^{\mathrm s},\mathbf a_n^{\mathrm s})$ is the fixed environment transition law from the within-round trajectory to the next system state, while $r_{i,n}^{\mathrm s}(\mathbf s_n^{\mathrm s},\mathbf a_n^{\mathrm s},\boldsymbol{\zeta}_n)$
is the corresponding reward calculation. Substituting \eqref{eq:fixed_fast_trajectory_distribution} into \eqref{eq:induced_slow_transition_reward} yields \eqref{eq:stagewise_stationarity}.
\end{IEEEproof}

Proposition~\ref{prop:stagewise_stationarity} explains the rationale for fixing the pretrained MAPPO policy during reallocation learning. If both policies were updated simultaneously, the same reallocation action from the same state could yield different expected outcomes as UAV control evolves. The reallocation policy would then need to track changes caused by both its own decisions and the evolving control policy, complicating value estimation and credit assignment. Stagewise training removes this source of variation, allowing graph-based IPPO to evaluate reallocation decisions under a consistent UAV control policy.

\subsection{Slow-Timescale Inter-Region Reallocation}
\label{subsec:slow_timescale_scheduling}

\begin{figure}[!t]
\centering
\includegraphics[width=\columnwidth]{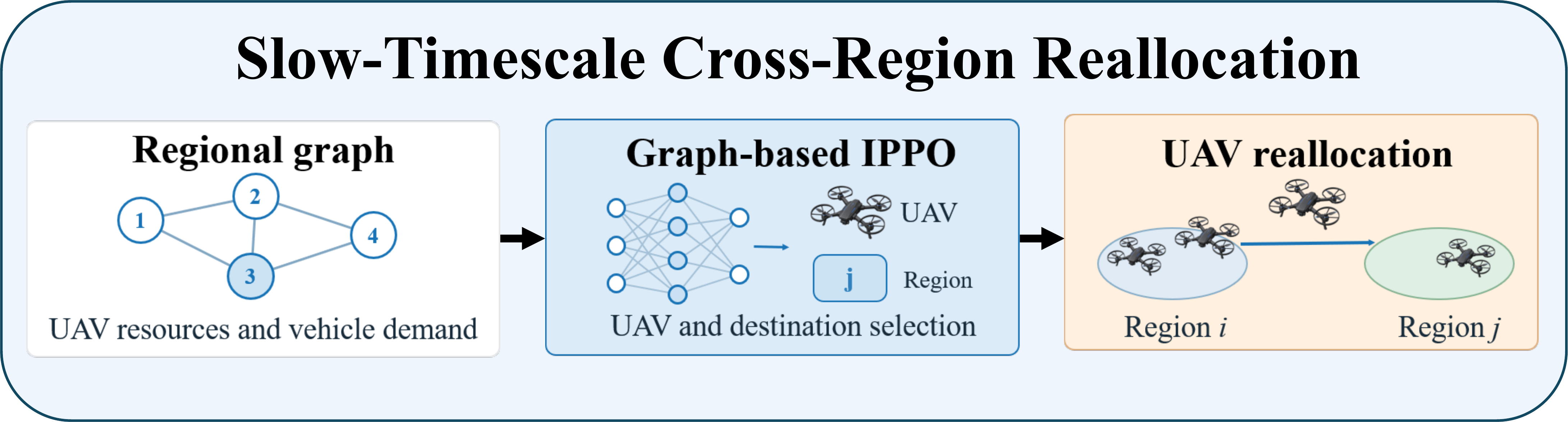}
\caption{Slow-timescale inter-region UAV reallocation using graph-based IPPO.}
\label{fig:slow_timescale_reallocation}
\end{figure}

The slow-timescale reallocation is modeled as a partially observable Markov decision process (POMDP), with each region acting as an agent, as illustrated in Fig.~\ref{fig:slow_timescale_reallocation}. Since dispatch decisions depend on both regional resource and demand conditions and feasible inter-region connections, we represent the system as a regional graph. At the beginning of slot $l=(n-1)K+1$, the regional graph for round $n$ is constructed as $\mathcal G_n=(\mathcal M,\mathcal E,\mathbf X_n,\mathbf Z)$, where $\mathcal E=\{(i,j)\mid j\in\mathcal N_{\bar i},\ i,j\in\mathcal M\}$ defines neighboring-region connections, $\mathbf X_n=\{\mathbf x_{i,n}\}_{i\in\mathcal M}$ collects regional node inputs, and $\mathbf Z$ is the edge-feature tensor. The node input of region $i$ is
\begin{equation}
\resizebox{0.89\linewidth}{!}{$\displaystyle
\mathbf x_{i,n}=\big(Q_i(l),V_i^{\mathrm{veh}}(l),U_i^{\mathrm a}(l),L_{i,n}^{\mathrm{c,min}},\mathbf D_{i,n}^{\mathrm{all}},\mathbf D_{i,n}^{\mathrm{uns}},\mathcal H_{i,n}\big).
$}
\label{eq:slow_node_feature}
\end{equation}
Here, $V_i^{\mathrm{veh}}(l)$ counts vehicles, while $U_i^{\mathrm a}(l)$ counts assigned UAVs, including those in transit toward region $i$. The minimum signed QoS-removal loss $L_{i,n}^{\mathrm{c,min}}$ is computed over available UAV candidates before sensing screening and set to zero when the source-reserve constraint precludes removal. The $4\times4$ maps $\mathbf D_{i,n}^{\mathrm{all}}$ and $\mathbf D_{i,n}^{\mathrm{uns}}$ give the densities of all vehicles and unsatisfied vehicles, respectively, relative to the regional center. The set $\mathcal H_{i,n}$ contains each assigned UAV's position, served-vehicle count, signed QoS-removal loss, transit status, sensing-removal loss, and identity. The edge feature $\mathbf z_{ij}=[d_{ij}^{\mathrm c},c_{j,x}-c_{i,x},c_{j,y}-c_{i,y}]^T$ contains the center distance $d_{ij}^{\mathrm c}=\|\mathbf c_j-\mathbf c_i\|$ and signed horizontal displacements.

Graph-based IPPO agents observe their own region and one-hop neighbors through $\mathbf o_{i,n}^{\mathrm s}$, including node inputs, edge features, and feasibility masks. A multilayer perceptron (MLP) and a convolutional network encode regional scalars and density maps, respectively. A shared UAV encoder processes $\mathcal H_{i,n}$; region-wise mean pooling and fusion with regional encodings produce fixed-dimensional node embeddings $\mathbf h_{i,n}$. A single four-head edge-aware graph attention network v2 (GATv2) layer~\cite{Brody2022GATv2} updates the local graph. For visible adjacent nodes $i$ and $j$, head $q$ computes the attention score
\begin{equation}
e_{ij,n}^{(q)}=\mathbf a_q^T\operatorname{LeakyReLU}\big(
\mathbf W_q^{\mathrm s}\mathbf h_{i,n}
+\mathbf W_q^{\mathrm t}\mathbf h_{j,n}
+\mathbf W_q^{\mathrm e}\mathbf z_{ij}\big),
\label{eq:slow_gat_attention}
\end{equation}
where LeakyReLU denotes the leaky rectified linear unit, and $\mathbf a_q$, $\mathbf W_q^{\mathrm s}$, $\mathbf W_q^{\mathrm t}$, and $\mathbf W_q^{\mathrm e}$ are trainable parameters. Each head aggregates messages $\mathbf W_q^{\mathrm t}\mathbf h_{j,n}+\mathbf W_q^{\mathrm e}\mathbf z_{ij}$ using softmax weights over adjacent nodes and the self-loop. The combined head outputs undergo residual addition and layer normalization. Actor and critic use separate encoders, shared across regions.

For each candidate neighboring destination, the actor scores eligible source UAVs using their embeddings, GATv2-updated source and destination embeddings, and edge features. Their softmax-weighted embedding sum, together with regional embeddings and edge features, determines the destination score. The actor first selects self-stay or a feasible destination, then selects a UAV conditioned on that destination if dispatching. Let $a_{i,n}^{\mathrm s}=(j,u)$ denote the action of dispatching UAV $u$ from region $i$ to region $j$ in round $n$. Its probability factorizes as
\begin{equation}
\pi_{\theta_{\mathrm s}}(a_{i,n}^{\mathrm s}\mid\mathbf o_{i,n}^{\mathrm s})
=
\pi_{\theta_{\mathrm s}}^{\mathrm d}(j\mid\mathbf o_{i,n}^{\mathrm s})
\pi_{\theta_{\mathrm s}}^{\mathrm{uav}}(u\mid j,\mathbf o_{i,n}^{\mathrm s}).
\label{eq:slow_joint_action}
\end{equation}
Feasibility masks exclude ineligible UAVs and conflicting destinations, enforcing self-stay when no dispatch is feasible. The local critic evaluates the source representation and feasible source-destination pairs from the same observation. Accepted actions form $\mathbf F_n$ and $\mathbf u_n$.

Since reallocation affects service through subsequent UAV operation, the slow-timescale reward is evaluated after executing $(\mathbf F_n,\mathbf u_n)$ and the associated fast-timescale control segment. A paired All-Stay rollout starts from the same state as the policy rollout at the beginning of the reallocation episode. It retains the initial regional allocation by making no cross-region dispatch while applying the fixed fast-timescale policy. This reference is maintained across rounds and shares the policy rollout's exogenous randomness. Let $Q_n^{\mathrm{AS}}$ and $J_n^{\mathrm{AS}}$ denote its QoS and fairness over the same round and evaluation window as $Q_n$ and $J_n$ in \eqref{eq:slow_timescale_performance}. All regional agents receive the same reward $r_n^{\mathrm s}$ at reallocation round $n$, defined as
\begin{equation}
r_n^{\mathrm s}=a_1(Q_n-Q_n^{\mathrm{AS}})+a_2(J_n-J_n^{\mathrm{AS}})-a_3E_n^{\mathrm d}.
\label{eq:slow_reward}
\end{equation}
This reward evaluates the collective outcome of dispatch and self-stay decisions while accounting for dispatch energy. The shared policy maximizes the expected discounted team return $\mathbb E_{\pi_{\theta_{\mathrm s}}}[\sum_{n=1}^{T_{\mathrm s}}\gamma_{\mathrm s}^{n-1}r_n^{\mathrm s}]$, where $\gamma_{\mathrm s}\in(0,1)$ is the discount factor.

\subsection{Fast-Timescale Intra-Region Control}
\label{subsec:fast_timescale_marl}

\begin{figure}[!t]
\centering
\includegraphics[width=\columnwidth]{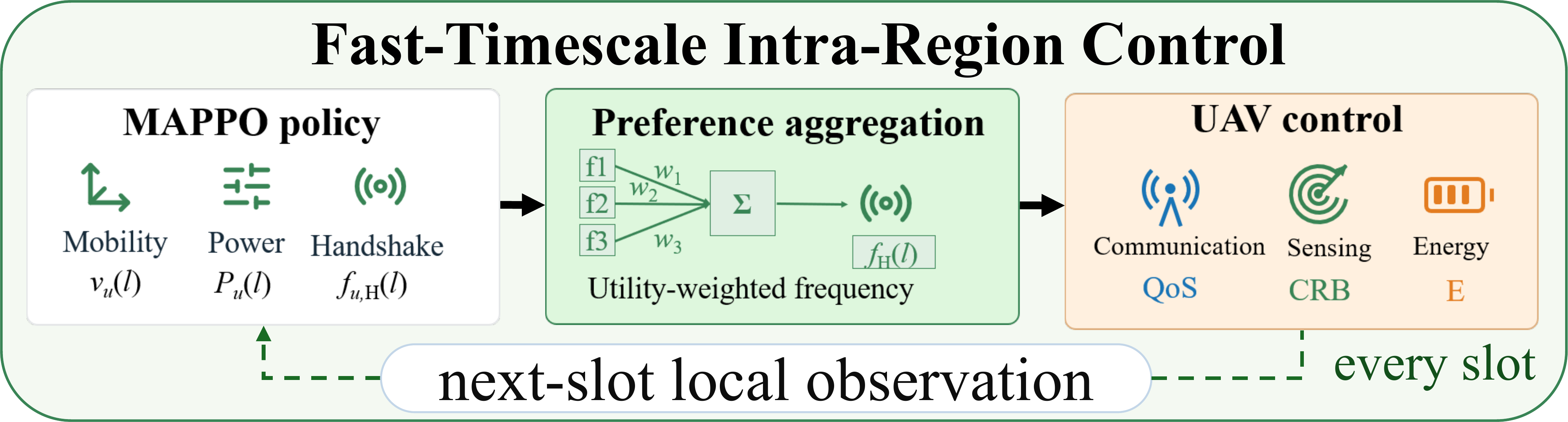}
\caption{Fast-timescale intra-region UAV control using MAPPO.}
\label{fig:fast_timescale_control}
\end{figure}

At the fast timescale, each service region acts as a local decision-making unit, as illustrated in Fig.~\ref{fig:fast_timescale_control}. UAVs in $\mathcal U_m(l)$ act as cooperative agents for local vehicle service and, when required, cross-region cooperative sensing. Given the slow-timescale UAV allocation, the decision process of region $m$ is modeled as a POMDP under centralized training with decentralized execution (CTDE), $\big(\mathcal S_m^{\mathrm f},\mathcal O_m^{\mathrm f},\mathcal A_m^{\mathrm f},\mathcal R_m^{\mathrm f},\mathcal P_m^{\mathrm f},\gamma_{\mathrm f}\big)$. Here, $\mathcal S_m^{\mathrm f}$ is the centralized state space, $\mathcal O_m^{\mathrm f}$ is the local observation space, $\mathcal A_m^{\mathrm f}$ is the continuous action space, $\mathcal R_m^{\mathrm f}$ denotes the regional reward function, $\mathcal P_m^{\mathrm f}$ denotes the state-transition law, and $\gamma_{\mathrm f}\in(0,1)$ is the discount factor.

For region $m$, the centralized state is
\begin{equation}
\resizebox{0.89\linewidth}{!}{$\displaystyle
\mathbf s_m^{\mathrm f}(l)=\big\{
\mathbf s_m^{\mathrm{uav}}(l),
\mathbf s_m^{\mathrm{nbr}}(l),
\mathbf s_m^{\mathrm{veh}}(l),
\mathbf s_m^{\mathrm{tar}}(l),
\mathbf s_m^{\mathrm{sync}}(l)
\big\}\in\mathcal S_m^{\mathrm f}.
$}
\label{eq:fast_state}
\end{equation}
Here, $\mathbf s_m^{\mathrm{uav}}(l)$ contains regional UAV positions and transmit powers, $\mathbf s_m^{\mathrm{nbr}}(l)$ contains neighboring-region UAV positions, and $\mathbf s_m^{\mathrm{veh}}(l)$ describes the regional vehicle distribution. The component $\mathbf s_m^{\mathrm{tar}}(l)$ contains available target information, while $\mathbf s_m^{\mathrm{sync}}(l)$ contains the last executed handshaking frequency $f_{\mathrm H}(l-1)$, the current sensing participation indicator $\alpha_m(l)$, and the latest sensing feedback $U_{\mathrm s}(l-1)$ and $U_{\mathrm s}^{m}(l-1)$. At $l=0$, all state quantities required for the first decision are initialized from the prescribed initial handshaking setting, UAV deployment, target state, and synchronization condition.

For UAV $u\in\mathcal U_m(l)$, the actor uses the local observation
\begin{equation}
\resizebox{0.89\linewidth}{!}{$\displaystyle
\mathbf o_u^{\mathrm f}(l)=\big\{
\mathbf o_u^{\mathrm{self}}(l),
\mathbf o_u^{\mathrm{veh}}(l),
\mathbf o_u^{\mathrm{nbr}}(l),
\mathbf o_u^{\mathrm{tar}}(l),
\mathbf o_u^{\mathrm{fb}}(l)
\big\}\in\mathcal O_m^{\mathrm f}.
$}
\label{eq:fast_observation}
\end{equation}
Here, $\mathbf o_u^{\mathrm{self}}(l)$ contains the position, velocity, transmit power, and individual communication-service status of UAV $u$. The component $\mathbf o_u^{\mathrm{veh}}(l)$ describes the distribution of locally reachable vehicles, while $\mathbf o_u^{\mathrm{nbr}}(l)$ contains the relative positions of nearby UAVs in the current and neighboring regions. The components $\mathbf o_u^{\mathrm{tar}}(l)$ and $\mathbf o_u^{\mathrm{fb}}(l)$ contain the relative target observation and available communication, sensing, and handshaking feedback, respectively. The observation also includes the regional hotspot position and a stable UAV identifier. Target- and handshaking-related entries are masked in regions not participating in cooperative sensing. From this observation, UAV $u$ produces the continuous action
\begin{equation}
\resizebox{0.89\linewidth}{!}{$\displaystyle
\mathbf a_u^{\mathrm f}(l)
=
\left[
v_{u,x}(l),
v_{u,y}(l),
v_{u,z}(l),
f_{u,\mathrm H}(l),
P_u(l)
\right]^T
\in\mathcal A_m^{\mathrm f},
$}
\label{eq:fast_action}
\end{equation}
where the first three components form the velocity $\mathbf v_u(l)$ and $P_u(l)$ denotes the transmit power. The handshaking output $f_{u,\mathrm H}(l)$ is a local preference that must be aggregated with those of other UAVs to obtain the frequency for execution. Since current sensing feedback is available only after execution, aggregation uses the previous slot's regional marginal sensing utilities. Preferences within region $i$ are first averaged as $\bar f_{i,\mathrm H}(l)=|\mathcal U_i(l)|^{-1}\sum_{u\in\mathcal U_i(l)}f_{u,\mathrm H}(l)$. The common handshaking frequency is then given by
\begin{equation}
f_{\mathrm H}(l)
=
\frac{
\sum\limits_{i\in\mathcal N_m}
\left[
U_{\mathrm s}^{i}(l-1)
\right]
\bar f_{i,\mathrm H}(l)
}{
\sum\limits_{i\in\mathcal N_m}
U_{\mathrm s}^{i}(l-1)+\epsilon_{\mathrm H}
},
\label{eq:common_handshaking_frequency}
\end{equation}
where $\epsilon_{\mathrm H}>0$ prevents a zero denominator. Regions with larger marginal sensing contributions receive greater weight. After execution, the sensing utility and regional marginal utilities are reevaluated to provide feedback for the next slot.

The reward $r_m^{\mathrm f}(l)$ of region $m$ at slot $l$ combines communication, sensing, and energy terms of \textbf{P2} with soft penalties for regional-boundary and inter-UAV safety-distance violations:
\begin{equation}
r_m^{\mathrm f}(l)
=
b_1Q_m(l)
+
b_2\alpha_m(l)U_{\mathrm s}(l)
-
b_3E_m(l)
-
\Pi_m^{\mathrm f}(l),
\label{eq:fast_reward}
\end{equation}
where $b_1$, $b_2$, and $b_3$ are the weights adopted in \textbf{P2}. The indicator $\alpha_m(l)$ activates the sensing term only when region $m$ participates in cooperative sensing. The soft penalty is $\Pi_m^{\mathrm f}(l)=\lambda_{\mathrm b}\Pi_m^{\mathrm b}(l)+\lambda_{\mathrm d}\Pi_m^{\mathrm d}(l)$, where $\Pi_m^{\mathrm b}(l)$ and $\Pi_m^{\mathrm d}(l)$ penalize boundary and safety-distance violations, respectively, with weights $\lambda_{\mathrm b}$ and $\lambda_{\mathrm d}$. The shared UAV policy $\pi_{\theta_{\mathrm f}}$, parameterized by $\theta_{\mathrm f}$, is trained under CTDE to maximize the discounted regional return $\mathbb E_{\pi_{\theta_{\mathrm f}}}[\sum_{l=1}^{T_{\mathrm f}}\gamma_{\mathrm f}^{l-1}r_m^{\mathrm f}(l)]$.

\subsection{Hierarchical Policy Optimization}
\label{subsec:hierarchical_policy_optimization}

\begin{figure}[!t]
\centering
\includegraphics[width=\columnwidth]{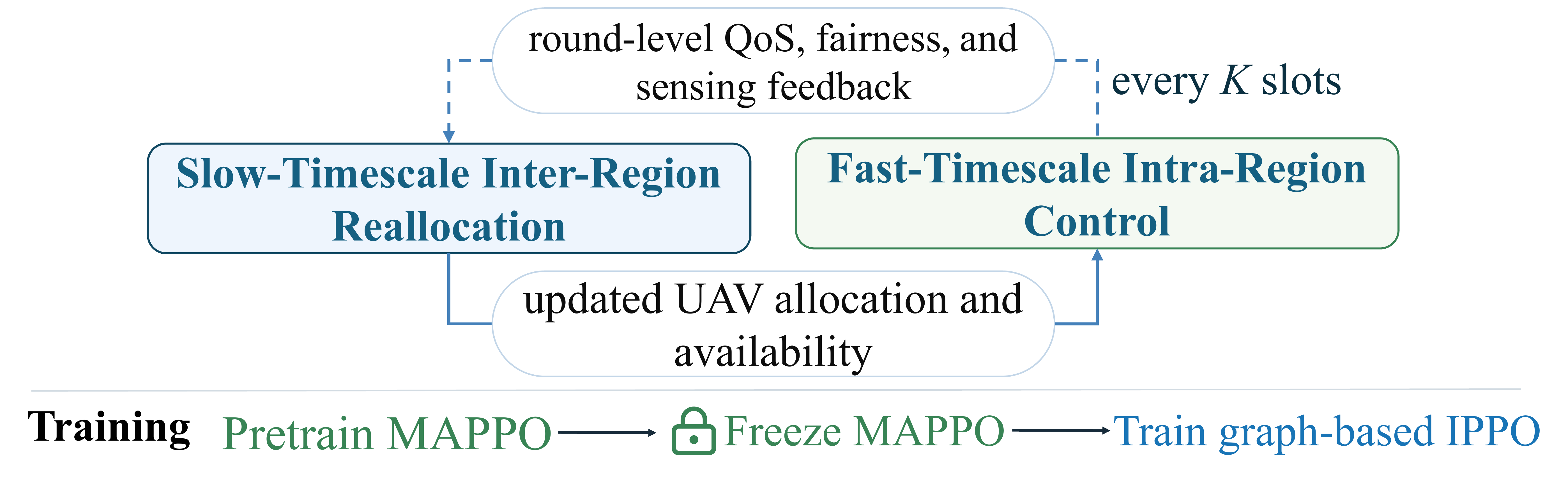}
\caption{Interaction between the two timescales and stagewise training of the hierarchical policies.}
\label{fig:hmarl_framework}
\end{figure}

Fig.~\ref{fig:hmarl_framework} summarizes the two-timescale interaction and stagewise training. Each fast-timescale pretraining episode includes one feasible cross-region UAV reallocation, with experience collected before and after reallocation to help the shared MAPPO policy adapt to allocation changes. Let $\pi_{\theta_{\mathrm f}}(\mathbf a_u^{\mathrm f}(l)|\mathbf o_u^{\mathrm f}(l))$ and $V_{\psi_{\mathrm f}}^{\mathrm f}(\mathbf s_m^{\mathrm f}(l))$ denote its action probability and centralized regional critic, respectively. The temporal-difference (TD) error and generalized advantage estimation (GAE) are computed as follows:
\begin{subequations}
\label{eq:fast_advantage_estimation}
\begin{align}
\delta_l^{\mathrm f}
&=
r_m^{\mathrm f}(l)
+
\gamma_{\mathrm f}
V_{\psi_{\mathrm f}}^{\mathrm f}
\big(\mathbf s_m^{\mathrm f}(l+1)\big)
-
V_{\psi_{\mathrm f}}^{\mathrm f}
\big(\mathbf s_m^{\mathrm f}(l)\big),
\label{eq:fast_td_error}
\\
\widehat A_l^{\mathrm f}
&=
\sum_{\ell=0}^{T_{\mathrm f}-l}
\left(\gamma_{\mathrm f}\lambda_{\mathrm f}\right)^\ell
\delta_{l+\ell}^{\mathrm f},
\label{eq:fast_gae}
\end{align}
\end{subequations}
where $\lambda_{\mathrm f}$ is the GAE parameter. The likelihood ratio of UAV $u$ is $\rho_{u,l}^{\mathrm f}(\theta_{\mathrm f})=\pi_{\theta_{\mathrm f}}(\mathbf a_u^{\mathrm f}(l)|\mathbf o_u^{\mathrm f}(l))/\pi_{\theta_{\mathrm f}^{\mathrm{old}}}(\mathbf a_u^{\mathrm f}(l)|\mathbf o_u^{\mathrm f}(l))$. The fast-timescale actor maximizes the clipped proximal policy optimization (PPO) objective
\begin{equation}
\begin{aligned}
\mathcal J_{\pi}^{\mathrm f}(\theta_{\mathrm f})
=
\mathbb E_{l,u}
\Big[
&
\min\Big(
\rho_{u,l}^{\mathrm f}\widehat A_l^{\mathrm f},
\operatorname{clip}\big(
\rho_{u,l}^{\mathrm f},
1-\epsilon_{\mathrm f},
1+\epsilon_{\mathrm f}
\big)
\widehat A_l^{\mathrm f}
\Big)
\\
&+
c_{\mathrm H}^{\mathrm f}
\mathcal H\!\left(
\pi_{\theta_{\mathrm f}}
(\cdot|\mathbf o_u^{\mathrm f}(l))
\right)
\Big],
\end{aligned}
\label{eq:fast_ppo_objective}
\end{equation}
where $\epsilon_{\mathrm f}$ is the clipping parameter, $c_{\mathrm H}^{\mathrm f}$ is the entropy coefficient, and $\mathcal H(\cdot)$ denotes policy entropy. The centralized critic is updated by minimizing
\begin{equation}
\mathcal L_V^{\mathrm f}(\psi_{\mathrm f})
=
\mathbb E_{l,m}
\left[
\ell_{\mathrm H}
\left(
V_{\psi_{\mathrm f}}^{\mathrm f}
(\mathbf s_m^{\mathrm f}(l))
-
\widehat R_l^{\mathrm f}
\right)
\right],
\label{eq:fast_critic_loss}
\end{equation}
where $\widehat R_l^{\mathrm f}$ is the normalized GAE-based return target and $\ell_{\mathrm H}(\cdot)$ denotes the Huber loss. With the pretrained fast-timescale policy fixed, graph-based IPPO updates regional actors and local critics $V_{\psi_{\mathrm s}}^{\mathrm s}(\mathbf o_{i,n}^{\mathrm s})$ from local rollouts, with their respective parameters shared across regions. During training, regions act sequentially in a random order under feasibility masks. The shared reward in~\eqref{eq:slow_reward} is evaluated after the associated $K$-slot control segment and used for GAE with discount factor $\gamma_{\mathrm s}$ and parameter $\lambda_{\mathrm s}^{\mathrm{GAE}}$. The clipped PPO actor objective uses the joint action probability in~\eqref{eq:slow_joint_action}, and the local critic fits normalized GAE return targets using the Huber loss in~\eqref{eq:fast_critic_loss}.

\begin{algorithm}[t]
\caption{Online Execution of H-MARL}
\label{alg:online_execution}
\begin{algorithmic}[1]
\REQUIRE Trained policies $\pi_{\theta_{\mathrm s}}, \pi_{\theta_{\mathrm f}}$; initial state and feedback; $K,T_{\mathrm f}$.
\FOR{$l=1,\ldots,T_{\mathrm f}$}
    \IF{$(l-1)\bmod K=0$}
        \STATE Set $n=(l-1)/K+1$; construct $\mathbf o_{i,n}^{\mathrm s}$ and feasibility masks.
        \FOR{each region $i$ in a fixed order}
            \STATE Select destination $j$ or self-stay by masked maximization of $\pi_{\theta_{\mathrm s}}^{\mathrm d}$; if $j\ne i$, similarly select $u$ using $\pi_{\theta_{\mathrm s}}^{\mathrm{uav}} (\cdot\mid j,\mathbf o_{i,n}^{\mathrm s})$. Update conflict masks.
        \ENDFOR
        \STATE Form and execute $(\mathbf F_n,\mathbf u_n)$.
    \ENDIF
    \STATE Construct $\mathbf o_u^{\mathrm f}(l)$ and obtain deterministic $\mathbf a_u^{\mathrm f}(l)$ using $\pi_{\theta_{\mathrm f}}$, with inactive components masked.
    \STATE Aggregate handshaking preferences into $f_{\mathrm H}(l)$.
    \STATE Execute $\mathbf v_u(l)$, $P_u(l)$, and $f_{\mathrm H}(l)$.
    \STATE Update the system state and communication, sensing, and energy feedback.
\ENDFOR
\end{algorithmic}
\end{algorithm}

Algorithm~\ref{alg:online_execution} summarizes online execution of the fixed policies. UAVs in transit suspend communication and sensing until arrival. For $U\ge M\ge1$, assume fixed network depths, attention-head counts, and map resolution. Let $d$ bound the input and hidden dimensions, including $U$-dependent UAV identity features. Network-wide communication and CRB evaluations cost $\mathcal O(C_Q)$ and $\mathcal O(C_B)$, respectively, where $C_Q=U^2+UV+V\log(1+V)$ and $C_B=U^2M^2+M^3$. The CRB bound follows from dense FIM construction and nuisance-parameter elimination with at most $U^2$ sensing paths and $\mathcal O(M)$ unknown parameters. The dense implementation processes $M$ local graphs, each padded to $M$ regional nodes with $U$ UAV slots per node. The fast-stage cost per slot is $C_{\mathrm f}=\mathcal O\!\bigl(Ud^2+U^2\log(1+U)+C_Q+MC_B\bigr)$, and the slow-stage cost per reallocation round is $C_{\mathrm s}=\mathcal O\!\bigl(M^2Ud^2+U(C_Q+C_B)\bigr)$.
These bounds include $\mathcal O(M)$ CRB evaluations per slot, up to $U$ candidate-removal evaluations reused across destinations per round, and $\mathcal O(U)$ energy accounting per slot.

\section{Simulations}
\label{sec:simulation_results}
In this section, we evaluate graph-based representation, inter-region UAV reallocation, intra-region UAV control, and cooperative sensing with adaptive handshaking, and examine the effects of reward weights.

\subsection{Simulation Setup}
We simulate ten UAVs in a $1000\,\mathrm{m}\times1000\,\mathrm{m}$ area divided into five Voronoi regions, labeled C0-C4 for $m=1,\ldots,5$. The initial regional vehicle counts are $[15,25,75,35,50]$, totaling 200 vehicles. Initial vehicle positions are Gaussian-distributed around regional hotspots with an $80~\mathrm{m}$ standard deviation per coordinate. They subsequently follow a random walk, with each horizontal velocity component independently sampled uniformly from $[-5,5]~\mathrm{m/s}$ at each slot. The target undergoes random motion, with each velocity component having a standard deviation of $\sigma_v=1~\mathrm{m/s}$. All regions are assumed to have the same residual phase-error variance floor, i.e., $\sigma_{i,0}^{2}=0.01~\mathrm{rad}^{2}$ for all $i\in\mathcal N_{\bar m}$. Table~\ref{tab:simulation_parameters} summarizes the main parameters. Fast-timescale training starts from a random UAV allocation with at least one UAV per region. Each episode has two control periods separated by one feasible random dispatch round. Slow-timescale training comprises five reallocation stages, including the initial allocation. The evaluated algorithms are summarized below:
\begin{itemize}
    \item \textit{IPPO}~\cite{Lin2025CEE}: This slow-timescale baseline uses fully connected actors and critics without graph-based representation, while retaining the PPO updates used by the proposed slow-timescale method.
    \item \textit{Independent advantage actor-critic (IA2C)}~\cite{Boggess2026Decentralized}: This slow-timescale baseline is the same as IPPO but applies independent advantage actor-critic updates.
    \item \textit{Demand-based}: This slow-timescale baseline prioritizes feasible neighboring regions with more unserved vehicles than the source and selects an eligible UAV minimizing the estimated source region QoS loss upon removal.
    \item \textit{All-Stay}: This baseline retains the initial regional UAV allocation without cross-region reallocation while applying the proposed fast-timescale method.
    \item \textit{Multi-agent deep deterministic policy gradient (MADDPG)}~\cite{Ji2026IACTS}: This fast-timescale baseline learns continuous UAV control using a shared deterministic actor and a centralized action-value critic with experience replay.
    \item \textit{Reward-based}~\cite{Liu2018RewardGreedy}: This fast-timescale baseline selects the joint UAV control action with the highest estimated instantaneous team reward from a predefined candidate set based on the latest available observations.
    \item \textit{Random}: At the slow timescale, this baseline randomly selects self-stay or a feasible neighbor and, if dispatching, an eligible UAV. At the fast timescale, it independently samples UAV velocity components, transmit powers, and handshaking preferences uniformly within bounds.
\end{itemize}

\begin{table}[!t]
\centering
\caption{Simulation Parameters}
\label{tab:simulation_parameters}
\footnotesize
\setlength{\tabcolsep}{2pt}
\renewcommand{\arraystretch}{1.08}
\begin{tabular}{p{0.59\columnwidth}p{0.33\columnwidth}}
\hline
Parameter & Value \\
\hline
Antenna half-power beamwidth $\Theta$ & $90^\circ$ \\
UAV altitude range $[z_{\min},z_{\max}]$ & $80$-$150~\mathrm{m}$ \\
Horizontal velocity component bound $V_{\mathrm{xy}}^{\max}$ & $15~\mathrm{m/s}$ \\
Vertical speed bound $V_{\mathrm z}^{\max}$ & $5~\mathrm{m/s}$ \\
Minimum UAV separation $d_{\min}$ & $15~\mathrm{m}$ \\
Carrier frequency $f_c$ & $2~\mathrm{GHz}$ \\
Bandwidth $B$ & $10~\mathrm{MHz}$ \\
Maximum transmit power $P_{\max}$ & $2~\mathrm{W}$ \\
Noise power over bandwidth $B$ & $10^{-13}~\mathrm{W}$ \\
Required vehicle rate $R_{\mathrm{th}}$ & $1~\mathrm{Mbit/s}$ \\
Handshaking frequency range $[f_{\mathrm H}^{\min},f_{\mathrm H}^{\max}]$ & $1$-$100~\mathrm{Hz}$ \\
\noalign{\vskip 2pt}
\hline
\end{tabular}
\end{table}

\begin{figure*}[!t]
\centering
\subfloat[Average reward]{\includegraphics[width=0.32\textwidth]{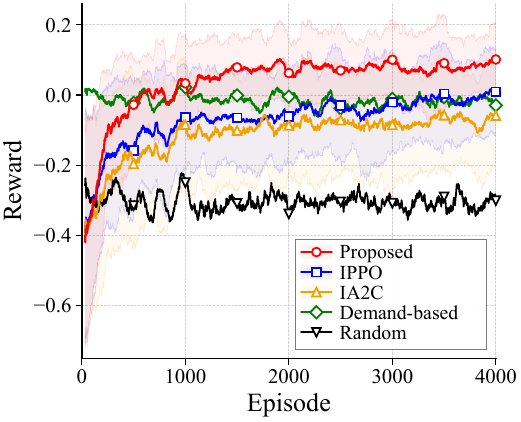}}\hfill
\subfloat[Communication QoS]{\includegraphics[width=0.32\textwidth]{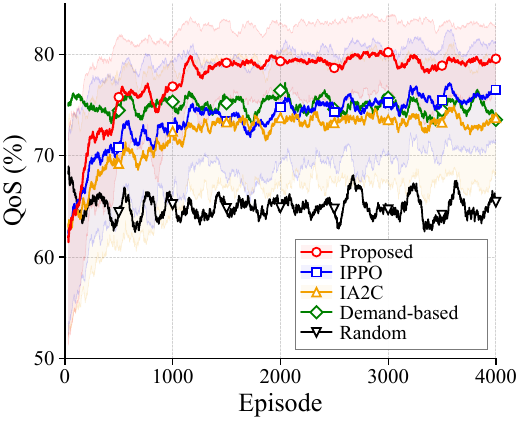}}\hfill
\subfloat[Regional fairness]{\includegraphics[width=0.32\textwidth]{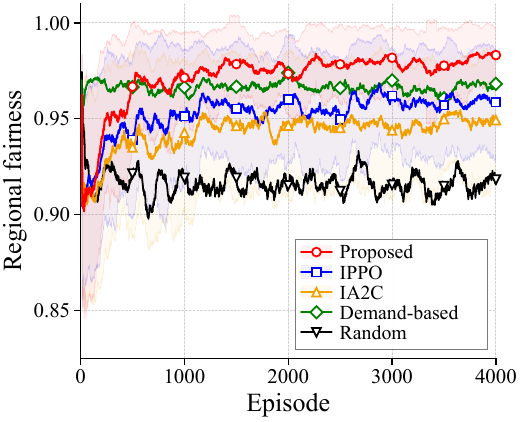}}
\caption{Training performance of different policies for inter-region UAV reallocation.}
\label{fig:slow_timescale_comparison}
\end{figure*}

\subsection{Graph-Based Representation Evaluation}
We compare the proposed slow-timescale method with IPPO to assess the contribution of graph-based representation, and include IA2C, Demand-based, and Random as additional baselines. Fig.~\ref{fig:slow_timescale_comparison} shows that the proposed slow-timescale method learns reallocation decisions that improve both communication service and regional fairness. Training curves show 100-episode moving averages, with shaded regions indicating within-window standard deviations for learning-based methods only. The proposed slow-timescale method achieves a QoS of $79.32\%$ and a regional fairness of 0.98, with the latter exceeding that of the comparison methods.

\subsection{Regional UAV Reallocation Assessment}
\begin{figure}[!t]
\centering
\includegraphics[width=0.8\columnwidth]{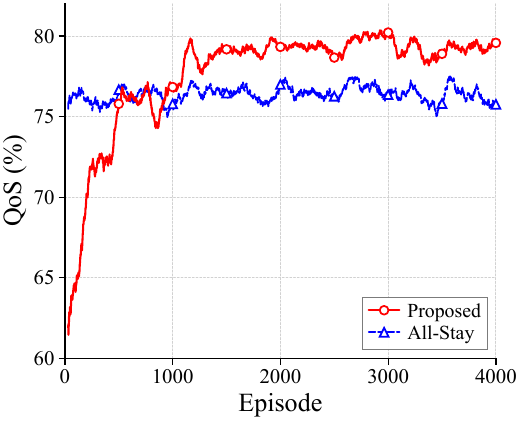}
\caption{Communication QoS with and without UAV reallocation.}
\label{fig:hierarchical_qos_comparison}
\end{figure}

Fig.~\ref{fig:hierarchical_qos_comparison} compares the proposed slow-timescale method with All-Stay to assess the benefit of slow-timescale reallocation. With the same fast-timescale policy, the proposed slow-timescale method achieves a mean QoS of $79.32\%$, versus $76.34\%$ for All-Stay, a gain of 2.98 percentage points.

\begin{table}[!t]
\centering
\caption{Regional UAV Reallocation Using the Proposed Slow-Timescale Method}
\label{tab:regional_reallocation}
\small
\setlength{\tabcolsep}{3pt}
\renewcommand{\arraystretch}{1.1}
\begin{tabular}{@{}ccccc@{}}
\hline
& & \multicolumn{3}{c}{UAVs: initial $\to$ final} \\
\cline{3-5}
\noalign{\vskip 2pt}
Region
& \shortstack{Final\\vehicle count}
& Overall
& \shortstack{High-gain\\group}
& \shortstack{Low-gain\\group} \\
\hline
C0 & 21.37 & $2.01\to2.06$ & $2.27\to2.06$ & $1.88\to2.08$ \\
C1 & 29.72 & $1.99\to1.63$ & $2.18\to1.57$ & $1.82\to1.67$ \\
C2 & 64.39 & $1.98\to2.35$ & $1.86\to2.40$ & $2.12\to2.29$ \\
C3 & 35.76 & $2.03\to1.98$ & $2.03\to1.95$ & $2.01\to2.00$ \\
C4 & 48.75 & $1.99\to1.98$ & $1.65\to2.01$ & $2.18\to1.96$ \\
\hline
\end{tabular}
\end{table}

Table~\ref{tab:regional_reallocation} summarizes UAV reallocation over the final 500 training episodes. Relative to All-Stay, the high-gain group achieves a QoS improvement above 5 percentage points and contains 124 episodes; the low-gain group achieves an improvement below 3 percentage points and contains 260 episodes. C2 has the largest final vehicle population and gains UAVs on average, whereas C1 releases UAVs. The larger low-gain group may partly reflect the random initialization: after assigning one UAV per region, the remaining UAVs are allocated with equal regional probabilities, favoring less concentrated allocations. This group starts with fewer UAVs in C1 and more in C2 and C4 than the high-gain group, leaving less scope to correct supply-demand mismatches through reallocation. Since the target is located in C0, the CRB constraint restricts UAV departures, so C0 does not exhibit the reduction in UAV numbers observed in C1 despite its smaller vehicle population. In the scenario with the largest QoS gain among these episodes, C1 initially has five UAVs, whereas C2 has only one UAV. Correcting this pronounced resource imbalance increases QoS from 57.94\% to 79.32\%.

\subsection{Intra-Region UAV Control Evaluation}
\begin{figure*}[!t]
\centering
\subfloat[Average reward]{\includegraphics[width=0.32\textwidth]{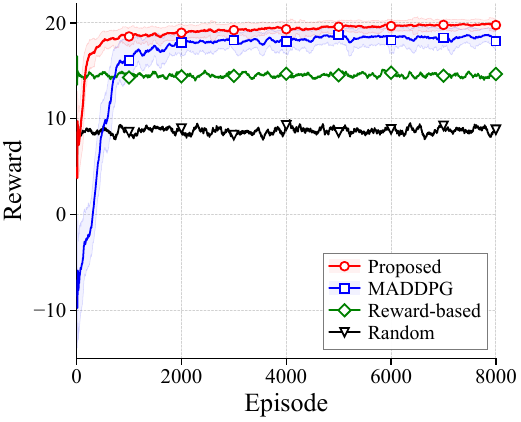}}\hfill
\subfloat[QoS]{\includegraphics[width=0.32\textwidth]{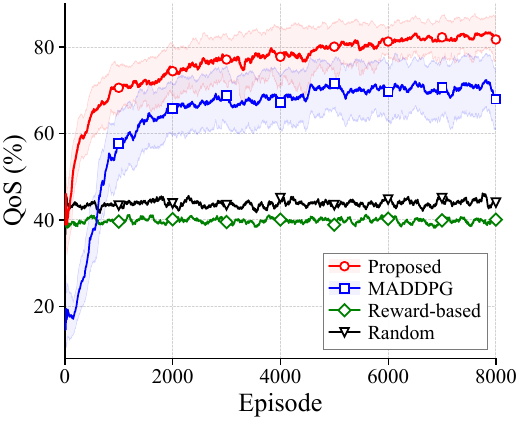}}\hfill
\subfloat[CRB]{\includegraphics[width=0.32\textwidth]{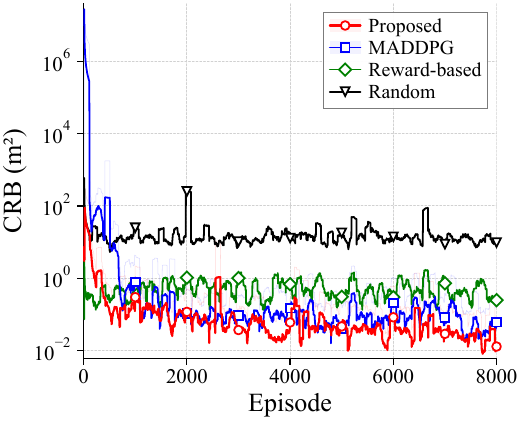}}
\caption{Training performance of different policies for intra-region UAV control.}
\label{fig:fast_timescale_comparison}
\end{figure*}

The proposed fast-timescale method is compared with MADDPG, Reward-based, and Random to assess its relative performance in terms of reward, communication QoS, and sensing CRB. As shown in Fig.~\ref{fig:fast_timescale_comparison}, the proposed fast-timescale method attains the highest mean reward of 19.83, with reward performance close to that of MADDPG. It achieves a QoS of $82.64\%$ and a mean CRB of $0.0237~\mathrm{m}^2$. Reward-based yields a lower CRB but poorer QoS than Random, indicating a tradeoff between sensing accuracy and communication service. This pattern is consistent with Reward-based favoring immediate estimated sensing gains without accounting for the longer-term consequences of its actions for vehicle coverage. It is worth noting that the initially concentrated vehicle distribution facilitates UAV coverage, whereas subsequent spatial dispersion can make coverage more challenging. Consequently, QoS may decline as slots progress even without cross-region UAV dispatch.

\subsection{Adaptive Handshaking and Cooperative Sensing}
We assess the contribution of adaptive handshaking by comparing the proposed mechanism with no handshaking and perfect synchronization using the same trajectories generated by the proposed fast-timescale method. The no-handshaking case removes synchronization prior information, while perfect synchronization provides an ideal reference with known regional phase offsets. A single-region case uses only sensing observations from UAVs in the target's region. As shown in Table~\ref{tab:handshaking}, adaptive handshaking reduces the mean CRB from $30.35~\mathrm{m}^2$ without handshaking to $0.0237~\mathrm{m}^2$, approaching the perfect-synchronization value of $0.0226~\mathrm{m}^2$. The remaining gap reflects residual inter-region phase uncertainty. The single-region case illustrates how limited observation diversity can compromise three-dimensional localization.

\begin{table}[!t]
\centering
\caption{Mean CRB under Different Sensing and Synchronization Conditions}
\label{tab:handshaking}
\footnotesize
\setlength{\tabcolsep}{4pt}
\renewcommand{\arraystretch}{1.12}
\begin{tabular}{lc}
\hline
Condition & Mean CRB ($\mathrm{m}^2$) \\
\hline
Adaptive handshaking & $0.0237$ \\
No handshaking & $30.35$ \\
Perfect synchronization & $0.0226$ \\
Single-region sensing\footnotemark & / \\
\hline
\end{tabular}
\end{table}
\footnotetext{In the considered scenario, single-region sensing may involve only one or two UAVs. Under the adopted sensing model, a finite three-dimensional localization CRB requires at least three UAVs with linearly independent viewing directions. Insufficient observation diversity can thus make the position FIM singular or ill-conditioned, resulting in anomalous CRB values.}

\begin{table}[!t]
\centering
\caption{Fast-Timescale Performance under Different Reward Weights}
\label{tab:weights_fast}
\footnotesize
\setlength{\tabcolsep}{2.5pt}
\renewcommand{\arraystretch}{1.12}
\begin{tabular*}{\columnwidth}{@{\extracolsep{\fill}}c*{3}{rr}}
\hline
\noalign{\vskip 2pt}
& \multicolumn{2}{c}{Proposed} & \multicolumn{2}{c}{MADDPG} & \multicolumn{2}{c}{Reward-based} \\
\noalign{\vskip 2pt}
\cline{2-7}
\noalign{\vskip 3pt}
$\beta$ & \multicolumn{1}{c}{\shortstack{QoS~$\uparrow$\\(\%)}} & \multicolumn{1}{c}{\shortstack{CRB~$\downarrow$\\($\mathrm{m}^2$)}} & \multicolumn{1}{c}{\shortstack{QoS~$\uparrow$\\(\%)}} & \multicolumn{1}{c}{\shortstack{CRB~$\downarrow$\\($\mathrm{m}^2$)}} & \multicolumn{1}{c}{\shortstack{QoS~$\uparrow$\\(\%)}} & \multicolumn{1}{c}{\shortstack{CRB~$\downarrow$\\($\mathrm{m}^2$)}} \\
\noalign{\vskip 2pt}
\hline
\noalign{\vskip 2pt}
$0.00$ & 41.70 & 0.0021 & 38.49 & 0.017 & 36.81 & 0.40 \\
$0.10$ & 83.23 & 0.0067 & 63.27 & 0.016 & 38.89 & 0.45 \\
$0.20$ & 83.32 & 0.0091 & 64.75 & 0.0084 & 38.93 & 0.50 \\
$0.30$ & 83.19 & 0.016 & 70.92 & 0.017 & 39.37 & 0.67 \\
$0.40$ & 84.02 & 0.020 & 64.92 & 0.61 & 39.97 & 0.56 \\
$0.55$ & 83.57 & 0.11 & 69.99 & 0.65 & 40.09 & 1.83 \\
$0.70$ & 83.26 & 0.41 & 64.94 & 0.53 & 42.04 & 4.55 \\
$0.85$ & 83.14 & 2.57 & 72.31 & 0.91 & 43.05 & 8.39 \\
$1.00$ & 83.82 & 5.53 & 73.41 & 1.22 & 43.91 & 25.34 \\
\noalign{\vskip 2pt}
\hline
\end{tabular*}
\end{table}

\begin{table}[!t]
\centering
\caption{Slow-Timescale Performance under Different Reward Weights}
\label{tab:weights_slow}
\footnotesize
\setlength{\tabcolsep}{2.5pt}
\renewcommand{\arraystretch}{1.12}
\begin{tabular*}{\columnwidth}{@{\extracolsep{\fill}}c*{3}{rr}}
\hline
\noalign{\vskip 2pt}
& \multicolumn{2}{c}{Proposed} & \multicolumn{2}{c}{IPPO} & \multicolumn{2}{c}{IA2C} \\
\noalign{\vskip 2pt}
\cline{2-7}
\noalign{\vskip 3pt}
$\xi$ & \multicolumn{1}{c}{\shortstack{QoS~$\uparrow$\\(\%)}} & \multicolumn{1}{c}{\shortstack{Fairness~$\uparrow$\\index}} & \multicolumn{1}{c}{\shortstack{QoS~$\uparrow$\\(\%)}} & \multicolumn{1}{c}{\shortstack{Fairness~$\uparrow$\\index}} & \multicolumn{1}{c}{\shortstack{QoS~$\uparrow$\\(\%)}} & \multicolumn{1}{c}{\shortstack{Fairness~$\uparrow$\\index}} \\
\noalign{\vskip 2pt}
\hline
\noalign{\vskip 2pt}
$0.00$ & 79.35 & 0.9888 & 74.95 & 0.9647 & 73.44 & 0.9657 \\
$0.25$ & 79.55 & 0.9878 & 77.76 & 0.9736 & 72.45 & 0.9623 \\
$0.50$ & 79.83 & 0.9876 & 78.79 & 0.9677 & 75.22 & 0.9620 \\
$0.63$ & 79.48 & 0.9842 & 79.15 & 0.9654 & 77.01 & 0.9569 \\
$0.75$ & 80.04 & 0.9863 & 79.92 & 0.9667 & 76.30 & 0.9526 \\
$0.80$ & 79.97 & 0.9798 & 78.66 & 0.9729 & 75.28 & 0.9433 \\
$0.86$ & 79.85 & 0.9827 & 78.77 & 0.9693 & 77.17 & 0.9588 \\
$0.93$ & 80.55 & 0.9801 & 77.41 & 0.9683 & 73.58 & 0.9551 \\
$1.00$ & 79.71 & 0.9810 & 77.28 & 0.9663 & 75.14 & 0.9453 \\
\noalign{\vskip 2pt}
\hline
\end{tabular*}
\end{table}

\subsection{Effects of Reward Weights}
We examine the effects of weighting communication QoS against CRB-based sensing performance in fast-timescale control, and communication QoS against Jain's fairness index in slow-timescale reallocation. Recall that $b_1$ and $b_2$ weight regional communication QoS and cooperative sensing utility, respectively, while $a_1$ and $a_2$ weight global communication QoS and regional fairness, respectively. The energy weights $b_3$ and $a_3$ are fixed. Keeping $b_1+b_2$ and $a_1+a_2$ constant, we vary the communication weight fractions $\beta=b_1/(b_1+b_2)$ and $\xi=a_1/(a_1+a_2)$, respectively. Table~\ref{tab:weights_fast} reports QoS and CRB for each fast-timescale weight setting. The proposed fast-timescale method achieves higher QoS than both baselines at every tested weight setting and a lower CRB in most cases. Its QoS initially increases with $\beta$ and then reaches a plateau. Once this plateau is reached, further increasing the communication weight yields little additional QoS improvement but markedly increases the CRB, reflecting the tradeoff between communication service and sensing accuracy. Table~\ref{tab:weights_slow} reports QoS and Jain's fairness index for each slow-timescale weight setting. The proposed slow-timescale method achieves higher QoS and fairness than both IPPO and IA2C at every tested weight setting. Appropriate emphasis on fairness may encourage UAV reallocation toward underserved regions, thereby improving both regional fairness and overall QoS. Weight selection therefore remains important for balancing communication QoS and service fairness across regions.

\section{Conclusions}
\label{sec:conclusion}
This paper proposed a cross-region cooperation framework for UAV swarms towards air-ground ISAC, in which UAV reallocation matches communication resources to vehicle demand and adaptive handshaking supports cooperative sensing across regions. A joint optimization problem was formulated for slow-timescale inter-region UAV reallocation and fast-timescale intra-region UAV control, accounting for communication, sensing, and energy consumption. An H-MARL approach was developed to solve the problem through coordinated regional UAV reallocation and local UAV control. Under the considered simulation settings, UAV reallocation improves QoS by over 20 percentage points in a scenario with severe resource-demand mismatches. The proposed reallocation policy achieves a regional fairness index of 0.98. Cooperative sensing with adaptive handshaking achieves a mean CRB of $0.0237~\mathrm{m}^2$, close to that under perfect synchronization.

\end{document}